\documentclass{aa}

\usepackage{soul}
\newcommand{\Ha}{\ifmmode {\mathrm{H}\alpha} \else H$\alpha$\fi\xspace}
\newcommand{\Hb}{\ifmmode {\mathrm{H}\beta} \else H$\beta$\fi\xspace}

\newcommand{\Hii}{\ifmmode \rm{H}\,\textsc{ii} \else H~{\textsc{ii}}\fi\xspace}
\newcommand{\Hi}{\ifmmode \rm{H}\,\textsc{i} \else H~{\textsc{i}}\fi\xspace}
\newcommand{\Nii}{\ifmmode [\text{N}\,\textsc{ii}]\lambda 6584 \else [N~{\scshape ii}]$\lambda 6584$\fi\xspace}
\newcommand{\nii}{\ifmmode [\text{N}\,\textsc{ii}] \else [N~{\scshape ii}]\fi\xspace}
\newcommand{\Oii}{\ifmmode [\rm{O}\,\textsc{ii}]\lambda 3727 \else [O~{\textsc{ii}}]$\lambda$3727\fi}
\newcommand{\oii}{\ifmmode [\rm{O}\,\textsc{ii}] \else [O~{\textsc{ii}}]\fi}
\newcommand{\Oiii}{\ifmmode [\rm{O}\,\textsc{iii}]\lambda 5007 \else [O~{\textsc{iii}}]$\lambda$5007\fi}
\newcommand{\oiii}{\ifmmode [\rm{O}\,\textsc{iii}] \else [O~{\textsc{iii}}]\fi}

\usepackage{orcidlink}
\usepackage{graphicx}
\usepackage{txfonts}
\begin{document}

   \title{Direct-method metallicity gradients derived from spectral stacking with SDSS-IV MaNGA}
   \titlerunning{Direct abundances from stacking using MaNGA}

   \author{Amir H. Khoram
          \inst{1,2,3}
          \orcidlink{https://orcid.org/0009-0009-6563-282X}
          \and
          Francesco Belfiore\inst{4}
          \orcidlink{https://orcid.org/0000-0002-2545-5752}
          }

   \institute{Dipartimento di Fisica e Astronomia, Università di Bologna, Via Gobetti 93/2, I-40129, Bologna, Italy
         \and
         INAF, Astrophysics and Space Science Observatory Bologna, Via P. Gobetti 93/3, I-40129 Bologna, Italy
        \and 
            INAF-Osservatorio Astronomico di Padova,
            vicolo dell’Osservatorio 5, 35122 Padova, Italy          
        \and 
             INAF - Osservatorio Astrofisico di Arcetri, Largo E. Fermi 5, 50125, Firenze, Italy\\
             }

   \date{}

\abstract{
Chemical abundances are key tracers of the cycle of baryons driving the evolution of galaxies. Most measurements of interstellar medium (ISM) abundance and metallicity gradients in galaxies are based, however, on model-dependent strong-line methods. Direct chemical abundances can be obtained via the detection of weak auroral lines, but such lines are too faint to be detected across large spectroscopic surveys of the local Universe. In this work we overcome this limitation and obtain metallicity gradients from direct method abundances by stacking spectra from the MaNGA integral field spectroscopy survey. In particular we stack 4140 star-forming galaxies across the star formation rate-stellar mass (SFR--$\mathrm{M_\star}$) plane and across six radial bins. 
We calculate electron temperatures for [OII], [SII], [NII], [SIII] and [OIII] across the majority of stacks. We find that the T[OII] $\sim$ T[SII] $\sim$ T[OII], as expected since these ions all trace the low-ionization zone of nebulae. The [OIII] temperatures become substantially larger than those of other ions at high metallicity, indicating potentially unaccounted for spectral contamination or additional physics. In light of this uncertainty we base our abundance calculation on the temperatures of [SIII] and the low-ionization ions.
We recover a mass-metallicity relation (MZR) similar to that obtained with different empirical calibrations. We do not find evidence, however, for a secondary dependence on SFR using direct metallicities. Finally, we derive  metallicity gradients that becomes steeper with stellar mass for $\log(M_\star/M_\odot) < 10.5$. At higher masses, the gradients flatten again, confirming with auroral line determinations the trends previously defined with strong-line calibrators.}  

\keywords{galaxies: abundances; galaxies: evolution; galaxies: general;
          galaxies: ISM; galaxies: gas content; ISM: abundances}

\maketitle

\section{Introduction}

Studying the abundance of heavy elements within the ISM of galaxies offers valuable insights into the physical mechanisms driving galaxy formation and evolution \citep{1980Tinsley,2012Rafelski}. The continuous inflow and outflow of baryonic matter within galaxies determine key galactic properties, linking M$_\star$, SFR, and metal content. Consequently, scaling relations such as the mass-metallicity relation \citep[MZR,][]{2004Tremonti,2008Kewley,2008Maiolino,2014Zahid, 2015Sanders,2020Jones,2023Nakajima} and the stellar mass-metallicity-star formation rate relation, also known as fundamental metallicity relation \citep[FMR,][]{2010Mannucci,2013Bothwell,2014Nakajima,2020Curti,2023Curti}, play pivotal roles as observational benchmarks in the development of galaxy evolution models.

Moreover, the chemical evolution of the ISM serves as a probe of the timescale for the assembly of galaxy discs \citep{1997Pagel,2010Spolaor,2012Rafelski,2019Maiolino&Mannucci}. Models describing galactic metallicity gradients, for example, generally support the inside-out scenario for the growth of galactic disks \citep{1976Larson,1989Matteucci&Francois,1999Boissier,2016Pezzulli}. Disagreements between the different models persist at higher redshift, where gradients are predicted to become shallower or steeper depending on the strength of feedback and the outflow properties \citep[e.g. ][]{2001Chiappini,2013Mott,2000Prantzos,2012Pilkington,2020Henriques,2021Sharda}.
Recent integral field spectroscopy (IFS) observations of nearby galaxies \citep[e.g.,][]{2011Rosales-Ortega,2012Croom,2014Sanchez,2015Ho,2015Bryant,2016Sanchez-Menguiano} have provided a comprehensive view of metallicity gradients in the local Universe, with extensions to high redshift being currently pursued with near-IR instruments, such as KMOS or NIRSpec on JWST.

These gradients have been derived using empirical calibrations or theoretical models, at the local Universe \citep[e.g.][]{2014Sanchez,2015Pilyugin,2017BelfioreO&N,2018Poetrodjojo,2020Berg,2020Mingozzi} and at higher redshifts \citep[e.g.][]{Curti2020, 2020Wang,2021Simons,2022Wang,2022Li,2024Venturi,2024Cheng}, with notably smaller sample sizes. They have demonstrated a general alignment with models grounded in the conventional inside-out framework of disk formation, which forecast a rapid self-enrichment process leading to elevated oxygen abundances and mostly negative metallicity gradients, especially when normalized to the optical size of the galaxy. 
Despite the good agreement between simulations and observations in the local universe, their divergent behavior at higher redshifts remains an open question. Simulations like IllustrisTNG and FIRE \citep[e.g.,][]{2017Ma,2021Hemler} as well as some observations \citep[e.g.,][]{2024Ju} show that at $\mathrm{z\gtrsim1}$ massive galaxies typically exhibit steeper negative metallicity gradients . However, several observations of high-redshift galaxies reveal a larger fraction of flat or inverted metallicity gradients compared to their local counterparts \citep[e.g.,][]{2020Curti,2021Simons,2024Cheng}.

The conflicting outcomes in low and high redshifts observed in certain earlier studies may be attributed to limited sample sizes and variations in the strong-line diagnostics employed to measure metallicity gradients. Utilizing diverse metallicity diagnostics can introduce substantial systematic errors \citep[see][]{2008Kewley,2008Ellison,2012López-Sánchez}. For instance, the observed flattening in metallicity gradients towards the central regions of the largest spiral galaxies, as discussed in the study by \cite{2016Zinchenko}, could be attributed to contamination from emission from other ionizing sources, as a low-ionization nuclear emission line regions (LINERs) found in massive galaxies \citep{Belfiore2016}, potentially affecting the reliability of strong line diagnostics. Alternatively, in extremely metal-rich central areas, metallicity levels may be approaching a saturation point, nearing their maximum attainable theoretical yield \citep[e.g.,][]{2017BelfioreO&N}. \\

Furthermore the relation between metallicity gradients and stellar mass of the galaxy, known as the mass--metallicity gradient relation (MZGR), can trace the disc assembly process \citep{2019Maiolino&Mannucci}. \cite{2014Sanchez} and \cite{2015Ho}  highlighted that normalizing radii to the galaxy effective radius ($R_e$) reduces the significant dispersion in metallicity gradients.

Using MaNGA (DR13) data, \cite{2017BelfioreO&N} found non--linear dependence of the metallicity gradient on stellar mass ($10^9-10^{11}$ M$_\odot$). \cite{2018Poetrodjojo} subsequently independently confirmed the changes in the metallicity gradient with stellar mass, using data from \citep[from SAMI Survey, ][]{2015Bryant} with stellar mass range $10^9-10^{10.5}$ M$_\odot$. These findings offered essential new constraints for chemical evolution models and their integration into hydrodynamical simulations, particularly regarding the balance between feedback strength and wind recycling in different mass regime.

Observationally, especially at sub-solar metallicities, the most precise approach for determining the oxygen abundance in the gas phase involves determining the effective temperature ($T_{\rm e}$) within \Hii regions, a method commonly referred to as the $T_{\rm e}$ technique or the `direct method' \citep{1992Pagel,2006Izotov}. This method hinges on the detection of faint auroral lines, which are not detected in individual objects in large spectroscopic surveys such as MaNGA or SAMI. 

The primary goal of this work is to directly measure the gas-phase metallicity gradients based on electron temperature extracted from auroral lines across a representative set of galaxies in the local Universe. We achieve this by stacking spectra of galaxies we expect to have similar metallicities. In particular, we stack the spectra of galaxies in bins in the M$_\star$ and SFR plane, since these two physical parameters are the main predictors of metallicity according to the fundamental metallicity relation (FMR) \citep[e.g.,][]{2010Mannucci,2013ApJ...765..140A,2014Nakajima}. 
We use data from the MaNGA survey \citep[DR17,][]{2022Abdurro'uf}, making our study the first of its kind to analyze the$\, \mathrm{T_e}  \, $metallicity gradient across the entire mass range of $10^{8.4}$ -- $10^{11.2} M_\odot$ .

The paper is organized as follows: In Sec. \ref{sec Data}, we present the galaxy sample obtained from the MaNGA survey. Sec. \ref{sec Data Analysis} outlines the methodology and our abundance determinations. Sec. \ref{sec result} presents the results, which are discussed in Sect. \ref{discussion}. Finally, Section  \ref{sec summary and conclusion} summarizes the conclusions of this study. We assume a $\mathrm{\Lambda CDM}$ cosmology with
$\mathrm{H_0 = 70\, km \,s^{-1} Mpc^{-1}\,, \Omega_m = 0.3, \,and \, \Omega_{\Lambda} = 0.7}$ .

\section{Data}\label{sec Data}

\begin{figure}

    \centering
    \includegraphics[width=\linewidth]{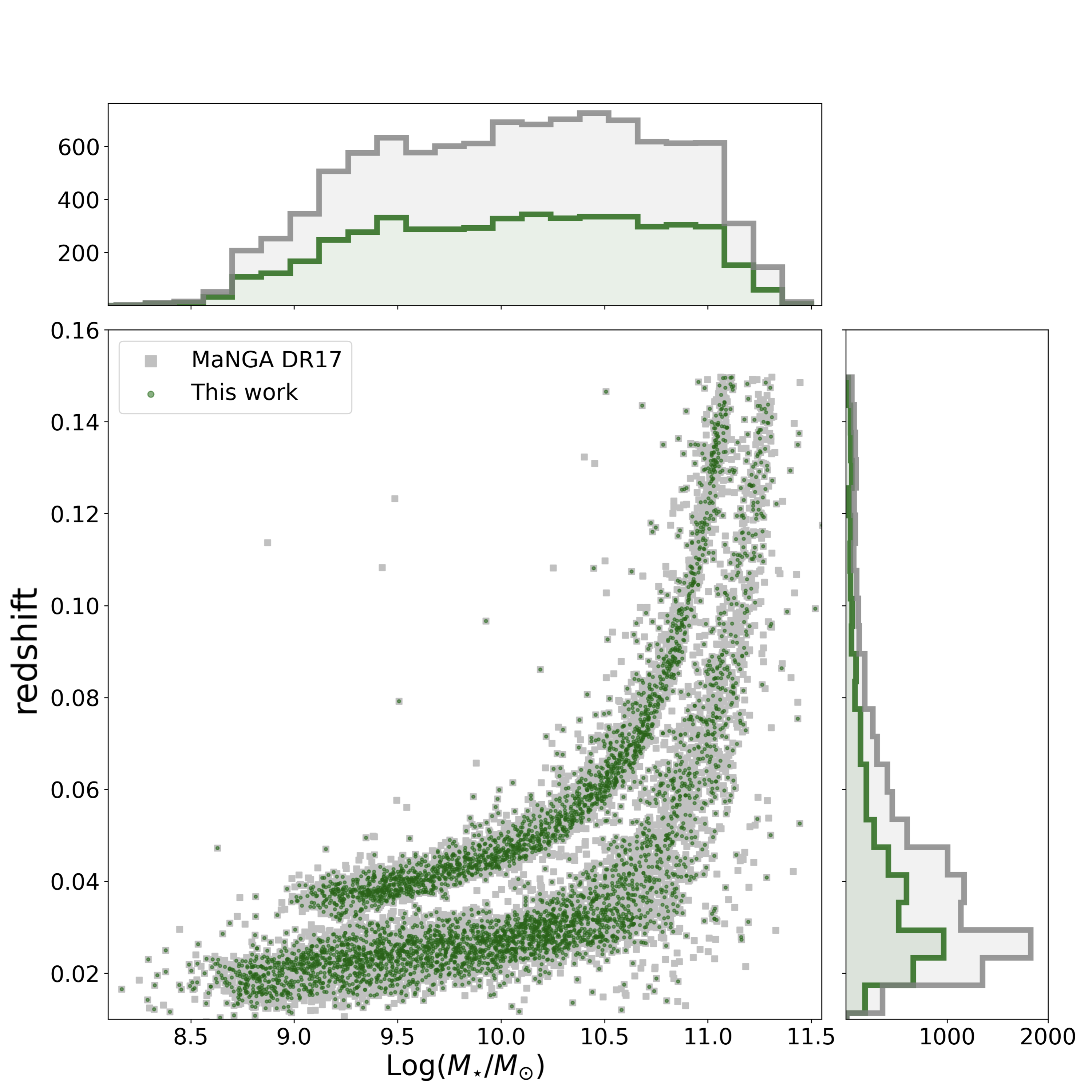}
    \caption{The redshift-$M_{\star}$ plane of the MaNGA sample utilized in this study. The star-forming galaxies (see Sect.\ref{Manga Data}) chosen for the study of gas-phase metallicity are shown in green and the entire MaNGA DR17 sample in gray. The redshift and stellar mass distributions for the selected and full sample used in this work are shown as green and grey histograms.}
    \label{fig: sample vs manga}
\end{figure}

\subsection{The MaNGA Data} \label{Manga Data}

\begin{figure*}[h]
    \centering
    \includegraphics[width=\linewidth]{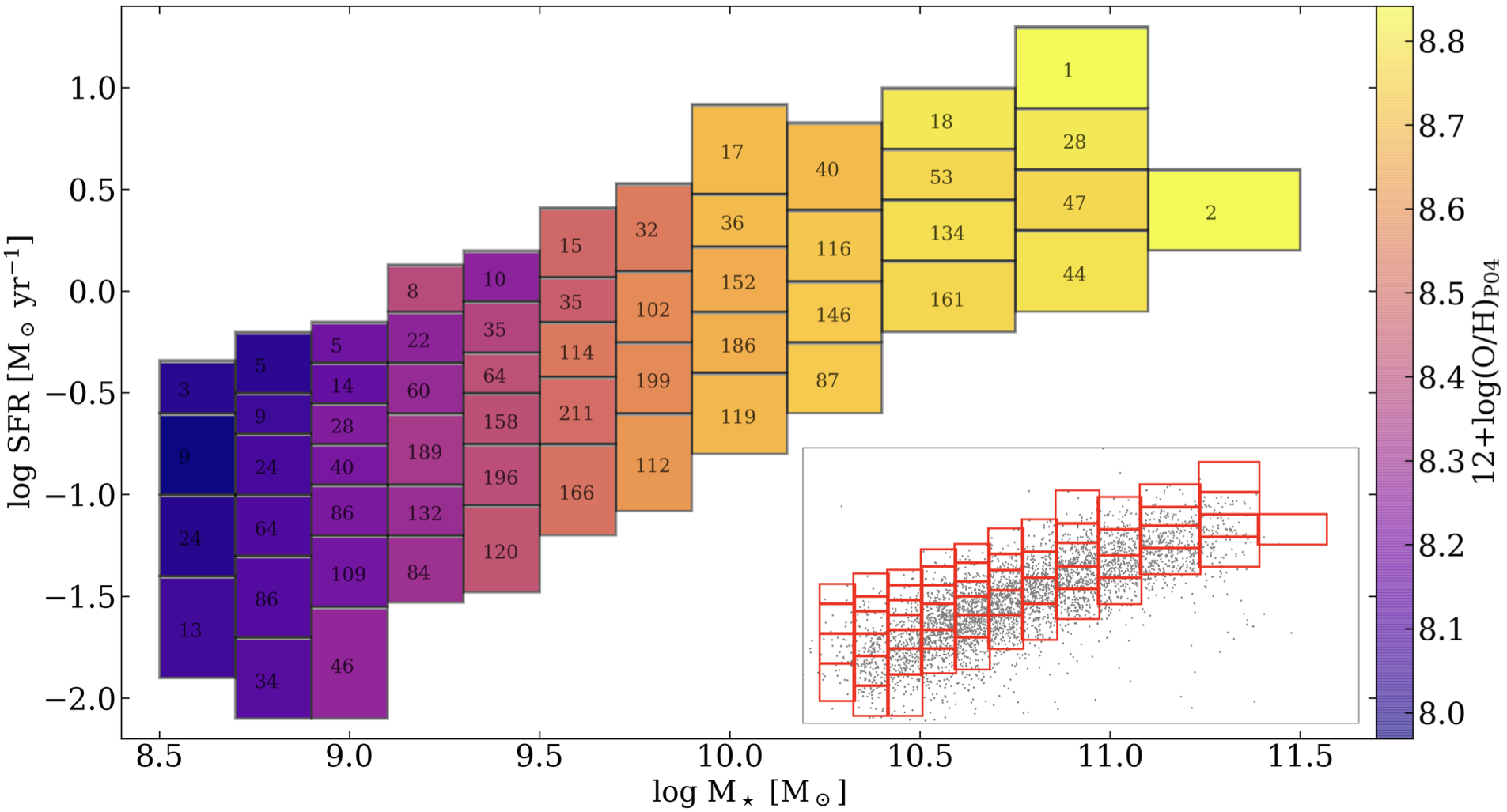}
    \caption{The position of the bins used for spectral stacking in this work in the SFR--$\mathrm{M_\star}$ plane. Each rectangle represents a bin, color-coded with its median strong-line metallicity using the calibration of \cite{2004MNRAS.348L..59P} and using extinction-corrected emission lines from the MaNGA \textsc{dap} catalog. The number of galaxies in each bin is reported. In the inset in the lower right our stacking grid is shown superimposed on the distribution of individual MaNGA star-forming galaxies in the SFR--$\mathrm{M_\star}$ plane. Note, integrated spectra, global and radially binned, are available online as introduced in Tab.\ref{tab:Global integrated spectra.} and \ref{tab:Radially binned integrated spectra}.}
    \label{fig:bins}
\end{figure*}

The fourth-generation Sloan Digital Sky Survey (SDSS) MaNGA (Mapping Nearby Galaxies at APO) project used integral field spectroscopy (IFS) to study the ionized ISM and stellar populations of a statistical sample of local galaxies at kpc scales. 
MaNGA data release 17 (DR17) includes observations of more than 10~000 local galaxies, in the redshift range $0.002 < \text{z} < 0.15$ (Fig.\ref{fig: sample vs manga}). MaNGA used integral field units composed of hexagonally packed fibre bundles of different sizes for scientific observations (from 19 to 127 fibres). In addition, the MaNGA instrument suite consisted of 92 single fibers for sky subtraction and a set of twelve 7-fiber mini-bundles for flux calibration \citep{2016AJ....151....8Y}. All fibers were fed into the dual beam BOSS spectrographs, which have a spectral resolution of R $\simeq 2000$ and span the wavelength range of 3600\AA \, to 10300\AA\,\citep{2013AJ....146...32S}.

MaNGA offers spatially resolved spectra that span a radial range up to 1.5 times the $R_e$ for the Primary sample, which accounts for approximately two-thirds of the total sample. For the Secondary sample, constituting roughly one-third of the total and located at slightly higher redshift, the spatially resolved spectra extend to 2.5 $R_e$ \citep{Law2015,Wake2017}. MaNGA DR17 data was reduced using the MaNGA reduction pipeline version v3\_1\_1, as detailed in \cite{2021Law,2016Law}. 

In this work we use the resulting reduced datacubes, with a pixel size of 0.5$''$. The typical point spread function (PSF) for the MaNGA datacubes is estimated to have a median full width at half-maximum (FWHM) of 2.5$''$. This corresponds to $\sim$ 1.5 kpc at the median distance of the MaNGA sample.

The MaNGA Data Analysis Pipeline \citep[\textsc{dap,}][]{2019AJ....158..160B,2019AJ....158..231W} performed spectral fitting for the determination of the stellar and gas kinematics, as well as the fluxes of various emission lines. The \textsc{dap} workflow employed adaptive Voronoi spatial binning \citep{2003Cappellari} to achieve a minimum target signal-to-noise ratio (SNR) of around 10 in the stellar continuum. After conducting measurements of stellar kinematics, it derived emission line fluxes using the full spectral fitting code pPXF \citep{2004CappellariandEmsellem,2017MNRAS.466..798C}.  For DR17 \textsc{dap} employed the MILES stellar library \citep{2006Sanchez} to extract stellar kinematics, while for modeling the stellar continuum within the emission line module it employed a subset of MaStar SSPs \citep{2020Maraston} derived from the MaStar stellar library \citep{2019Yan}. In this work, we employ the integrated emission line fluxes reported in \textsc{dap} v3\_1\_1 catalog to select a sample of star-forming galaxies from the full MaNGA sample. We also employ line fluxes and velocities from the emission lines maps released as part of DR17 to perform a selection of spaxels and to de-redshift the spectra before stacking. 

We adopt the integrated stellar masses estimates from the extended NSA targeting catalog \citep{2011Blanton}. These are obtained from fitting of the SDSS imaging data and adopt the \citealt{2003PASP..115..763C} initial mass function.
The global SFR for each galaxy is obtained from its dust-corrected H$\alpha$ luminosity, using the line fluxes from the integrated \textsc{dap} catalog, and adopting the conversion factor in \cite{2011ApJ...741..124H}. While discrepancies in the apertures used for stellar mass and SFR measurements may introduce some scatter,  \cite{Belfiore2018} found that aperture effects play a fairly minor role. In particular, they compared masses derived from within the MaNGA bundle and from integrated photometry and SFR measured within 1.5 $\rm~R_e$ and $2.5~R_e$, finding median offsets $<$ 0.05 dex.

Finally, we calculate deprojected galactocentric radii using the semi-axis ratio ($b/a$) from the MaNGA NSA catalog obtained via elliptical Petrosian analysis. Using the semi-axis ratio, galaxy inclination ($i$) is computed assuming constant oblateness $q = 0.13$ \citep{1994AJ....107.2036G} using 
\begin{equation}\label{eq:deprojection}
    \cos(i) = \frac{(b/a)^2 - q^2}{1 -  q^2}.
\end{equation}
The NSA elliptical Petrosian effective radius $R_e$ in the $r$-band is used as a normalizing scale-length for deriving gradients.

In order to measure metallicity gradients we select a sub-sample of star-forming galaxies from the full MaNGA dataset. In particular, we select galaxies that are classified as star-forming using the Baldwin, Phillips, and Terlevich (BPT) diagrams \citep{1981PASP...93....5B} using the fluxes extracted from their integrated spectra, as reported in MaNGA \textsc{dap} catalog.

Galaxies are classified as star-forming if they lie below the \cite{2003MNRAS.346.1055K} demarcation line criteria in the [O III]$\lambda$5007/H\(\beta\) vs. [N II]$\lambda$6583/H\(\alpha\)  diagram and the \cite{2001ApJ...556..121K} line in the [O III]$\lambda$5007/H\(\beta\) vs. [S II]$\lambda$$\lambda$6716,6731/H\(\alpha\) diagram.
These criteria select 4748 galaxies, spanning the stellar mass range $8.4<\log(\text{M}_\star/\text{M}_\odot)<11.5$. Note that this represents the number of star-forming galaxies in MaNGA sample, and while our binning captures the majority, not all of them are included in the $\mathrm{SFR-M_\star}$ bins (see Sect.\ref{stacking procedure}).

\subsection{Stacking Procedure} 
\label{stacking procedure}

We aim to measure weak auroral lines in spectral stacks.
Our approach consists in stacking in radial bins and across galaxy properties that select galaxies with similar metallicities. Assuming that a galaxy's integrated metallicity is described by the FMR, we bin galaxies across the M$_\star$ and SFR plane.

We define 56 bins in the SFR--$\mathrm{M_\star}$ plane (Fig.\ref{fig:bins}) , with 12 bins in stellar mass and a variable number (up to seven) bins in SFR. This binning scheme includes 4140 star-forming galaxies, excluding approximately 13\% of the sample, which lies outside these bins. The bins span $\log (\text{M}_{\star}/\text{M}_{\odot})$ = [8.4 -11.5] and  $\log(\rm SFR) \sim-2.0$ to $+1.0$ with an average continuum S/N of 37.5 at $\sim5000\AA$. The number of galaxies in each bin ranges from 1 to 211, with an average of 45 galaxies per bin. We tested various alternative binning configurations in this space, but found that the current one is the best compromise between the width of the bins and the ability to detect auroral lines. 

Before stacking the spectra we correct for their velocity shifts by using the MaNGA velocity map and the systemic redshift of the galaxy, shifting each spaxel's spectral data to its rest-frame wavelength.
We also apply an extinction correction to the stacked spectra of individual galaxies. This correction uses the extinction curve of \cite{1994O'Donnell}, assuming an intrinsic Balmer decrement of H$\alpha$/H$\beta$ = 2.86. The H$\alpha$ and H$\beta$ values are determined by modeling the stellar continuum and emission lines, as described in Sec.\ref{spectralfitting}. Consequently, we utilize the reddening-corrected spectra of individual galaxies within each SFR--$\mathrm{M_\star}$ bin to generate the final stacked (averaged) spectra for each respective bin.
To derive `integrated' abundances, spaxels are considered up to the radius covered by all samples, which is 1.5 $R_e$, and in the spatially resolved context, the analysis is conducted within different radial bins, as mentioned below.

Using extinction-corrected emission lines, the global strong-line metallicity of each galaxy was determined using the empirical calibration introduced by \cite{2004MNRAS.348L..59P}. The median metallicity of each bin, represented by the bin's color code in Fig. \ref{fig:bins}, reflects the average metallicity within that bin. The standard deviation of metallicity values in each bin is small, 0.07 dex on average.

 For spatially resolved analysis, we stack spectra of galaxies in different radial bins. Binning radially is a compromise between maintaining a sufficient number of spaxels to guarantee the detection of auroral lines and having a meaningful number of radial bins to determine the shape the gradient. For each bin in the $\rm M_\star$-SFR plane we draw a set of radial bins. We have experimented with equidistant radial bins, but eventually converged on a set of radial bins that more closely achieve a fixed signal-to-noise ratio. The bins we defined have boundaries at \(0.0, 0.35, 0.65, 0.85, 1.1, 1.5, \text{and } 2.5\) $\rm R_e$. Therefore, we construct six radial stacks for each bin in the SFR--$\mathrm{M_\star}$ plane.

 In our stacking analysis we exclusively incorporate spaxels that meet the star-forming classification criteria defined by the BPT method cuts discussed above for integrated galaxies. For spaxels, we use the line maps obtained using the MaNGA \textsc{DAP}. Spaxels with flagged H$\alpha$ velocity or flux measurements are omitted from the stacks.

\section{Data Analysis}\label{sec Data Analysis}

\subsection{Spectral Fitting}\label{spectralfitting}

We perform spectral fitting of the stacked spectra using pPXF \citep{2017MNRAS.466..798C}, and modelling the continuum with 32 E\-MILES simple stellar population templates \citep{2016MNRAS.463.3409V}. These models were generated using the initial mass function by \cite{2003PASP..115..763C}, BaSTI isochrones \citep{2004ApJ...612..168P}, eight ages ranging from 0.15 to 14 Gyr (logarithmically spaced in steps of 0.22 dex), four metallicities [Z/H]= [\-1.5, \-0.35, 0.06, 0.4]. During the fit we employ multiplicative polynomials of sixth degree.

\begin{table}
\caption{All measured emission lines in this study. Each box represents the wavelength window over which the spectra are modelled. Spectral line kinematics are tied to the most prominent line within each interval, denoted in bold.}
\renewcommand{\arraystretch}{1.15}
\resizebox{\columnwidth}{!}{%

\begin{tabular}{|lr|lr|}

\hline
 Atom/Ion   &   $\lambda$(\AA) & Atom/Ion   &   $\lambda$(\AA) \\
\hline

 OII          &   3726.03  &  \textbf{OIII}&   \textbf{5006.84}\\  
 \textbf{OII} &   \textbf{3728.73} &  HeI  &   5016.51\\  \cline{3-4}
 H9           &   3751.09 &  NII           &   5754.59 \\
 Hth          &   3798.96 &  \textbf{HeI}  &   \textbf{5875.61}\\  \cline{3-4}
 Ht           &   3836.26 &  OI            &   6300.30 \\
 NeIII        &   3868.69 &  SIII          &   6312.06\\
\cline{1-2} \cline{3-4}
 NeIII        &   3967.40 &  \textbf{OI}   &   \textbf{6363.78}\\ 
 He           &   3970.07 &  NII           &   6548.05\\
 SII          &   4068.60 &  \textbf{Ha}   &   \textbf{6562.82}\\
 SII          &   4076.35 &  NII           &   6583.45\\
 \textbf{Hd}  &   \textbf{4101.73} &  HeI           &   6678.15 \\
 FeII         &   4288.10 &  HeI  &   6680.05 \\ \cline{1-2}
 \textbf{Hg}  &   \textbf{4340.46} &  SII           &   6716.44 \\
 FeII         &   4359.33 &  SII  &   6730.81 \\ \cline{3-4}
 OIII         &   4363.20 & \textbf{ArIII} &   \textbf{7135.67} \\
 FeIII        &   4659.10 &  OII           &   7318.92 \\ \cline{1-2}
 HeII         &   4685.70 &  OII           &   7319.99 \\
 ArIV         &   4713.00 & OII           &   7329.67 \\
 \textbf{Hb}  &   \textbf{4861.33} & OII            &   7330.73 \\ \cline{3-4}
 OIII         &   4958.91 & TiII  &   8448.36 \\ 
 HeI          &   4987.44 & FeII           &   8574.85 \\ \cline{3-4}
              &           & \textbf{SIII}  &   \textbf{9068.61} \\

\hline
\end{tabular}
}

\label{tab:EmLine Table}
\end{table}

\begin{figure*}
    \centering
    \includegraphics[width=\linewidth,height=0.82\textheight,keepaspectratio]{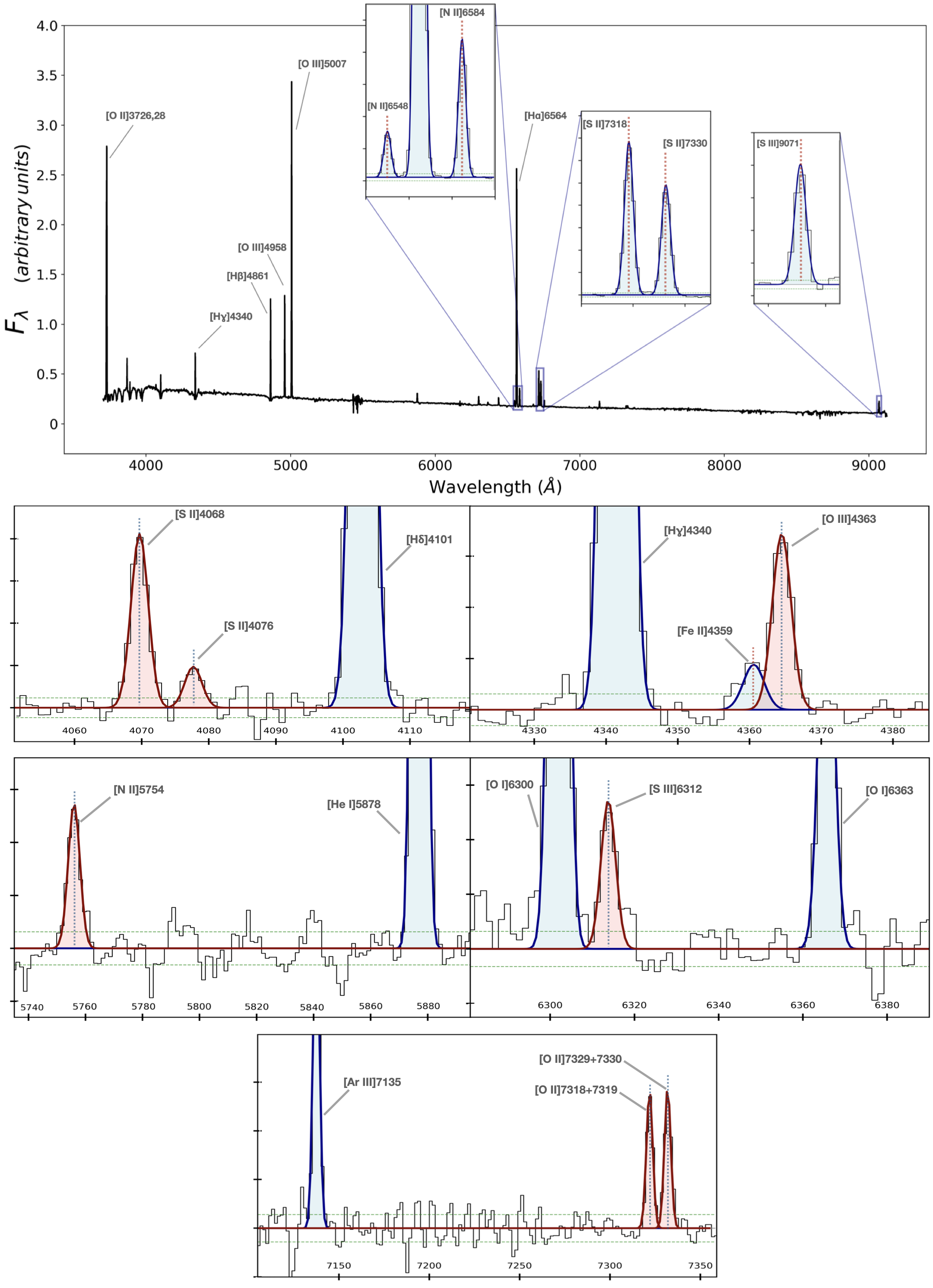}
    \caption{Topmost panel: The stacked spectrum of a SFR--$\mathrm{M_\star}$ bin ($\text{M}_\star = 10^{8.65} \, \text{M}_\odot$ and  $SFR = 1-^{-1.3} \, \text{M}_{\odot} \text{yr}^{-1}$) , showing zoom-ins of the spectral fits around some of the strong nebular lines. In the subplots below we zoom on the fit residuals in specific spectral regions to highlight the detection of faint, temperature-sensitive auroral lines (in red) and other nearby emission lines (blue). The green dashed lines represent the standard deviation of fits residual in each wavelength window. Note, emission line fluxes and associated uncertainties, global and radially binned, are available online as introduced in Tab.\ref{tab:Rad Flux} and \ref{tab:Global fluxes}.}
    \label{fig:full spectrum}
\end{figure*}

We model the spectrum in different spectral windows approximately $ 200$\AA\ wide, where the velocity dispersion of emission lines are tied to the strongest line's $\sigma_\textrm{v}$ in each window (Tab.\ref{tab:EmLine Table}). This procedure was found to produce smaller residuals than performing the pPXF fit over the entire wavelength range.
We impose a few additional constraints in the process of line fitting. In the case of the [OIII]$\lambda$4959,5007 and [NII]$\lambda$6548,6583 doublets, the dispersion of the individual components of the doublet is fixed and their amplitude ratios are set to the ratios of the relative Einstein coefficients. For [SIII] we only model the [SIII]$\lambda$9068 emission line since the [SIII]$\lambda$9530 line is either highly contaminated by the sky emission or for galaxies at the z$>0.08$, falls outside the MaNGA wavelength range. Furthermore, we mask all spectra in the range 5887\AA\, to 5903\AA\  to avoid any potential residuals from the strong sky line at $\lambda$5894.6\AA, which may overlap with the [N II]$\lambda$5754 emission line in the redshift range $0.022<   z <0.026$. As a result, the [NII]$\lambda$5754 is masked for $\sim$ 8\% of our galaxy sample. 

pPXF estimates its formal uncertainties from the covariance matrix of the standard errors in the fitted parameters, which may lead to underestimations in the parameter uncertainties. To provide a more realistic flux error we multiply the formal pPXF flux error by the ratio of the modeled residual standard deviation over the median flux error in each wavelength window. This process effectively rescales the flux uncertainties to match those of the fit residuals, therefore taking into account the errors introduced by imperfect continuum subtraction.

\subsection{Spectral contamination of the [OIII]$\lambda$4363 emission line}

We find at least one emission feature between 4358 \AA\ and 4362 \AA\ that is blended with the [O III]$\lambda$4363 auroral line, more commonly in the high metallicity stacks. Such contamination of the [OIII]$\lambda$4363 line has previously been reported in the literature \citep{2013ApJ...765..140A, 2017MNRAS.465.1384C}. Although the exact nature of this feature is unknown, it is at least in part constituted by the [Fe II]$\lambda$4360 emission line. 

\begin{figure}
    \centering
    \includegraphics[width=\linewidth]{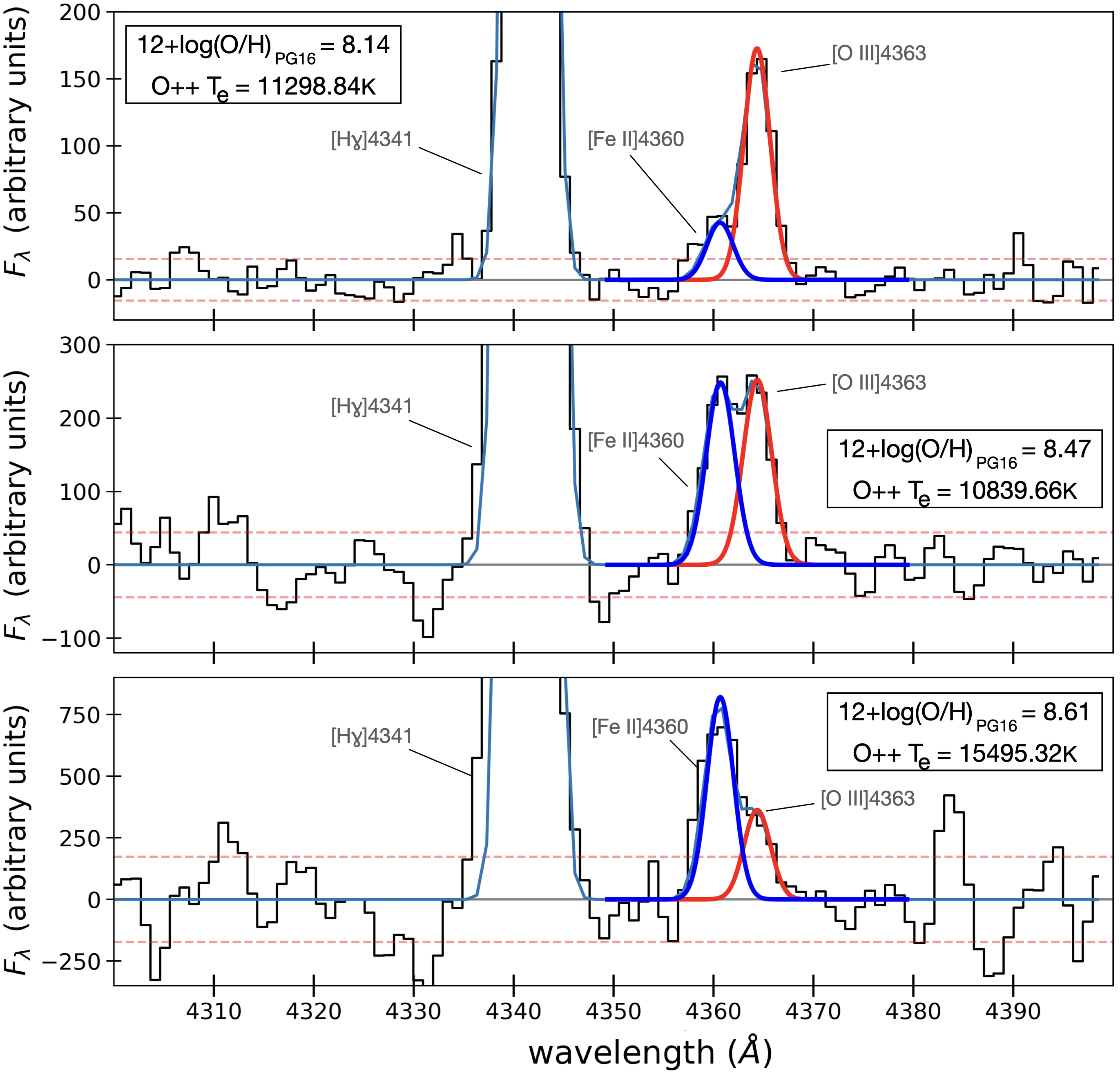}
    \caption{ Spectral residuals in the wavelength range near the [O III]$\lambda$4363 line, after stellar continuum subtraction. The red and blue Gaussian fits represent [O III]$\lambda$4363 and [Fe II]$\lambda$4360 emission lines respectively. The dashed lines illustrate the standard deviation of fits residual in the corresponding windows. The strong-line metallicity and measured [O III] temperature of each stack are reported in the panels. The contamination of the [O III]$\lambda$4363 line becomes more relevant with increasing metallicity. }
    \label{fig:O3_contamination}
\end{figure}

To mitigate the influence of [Fe II]$\lambda$4360 contamination during fitting, we set its flux to $0.73\times$ that of the nearby, isolated [Fe II]$\lambda$4288 line. This ratio follows that of the corresponding Einstein coefficient since both lines originate from the same atomic upper level. We also simultaneously fit the features around 4363\AA\ by linking their velocity widths and central wavelengths to \(\text{H}\gamma\).  The various components of the fit are shown in Fig. \ref{fig:O3_contamination} for three SFR--$\mathrm{M_\star}$ bins with different strong-line metallicities.

\subsection{Density and Temperature Measurements}\label{temp and density section}

\begin{figure*}
    \centering
    \includegraphics[width=\linewidth]{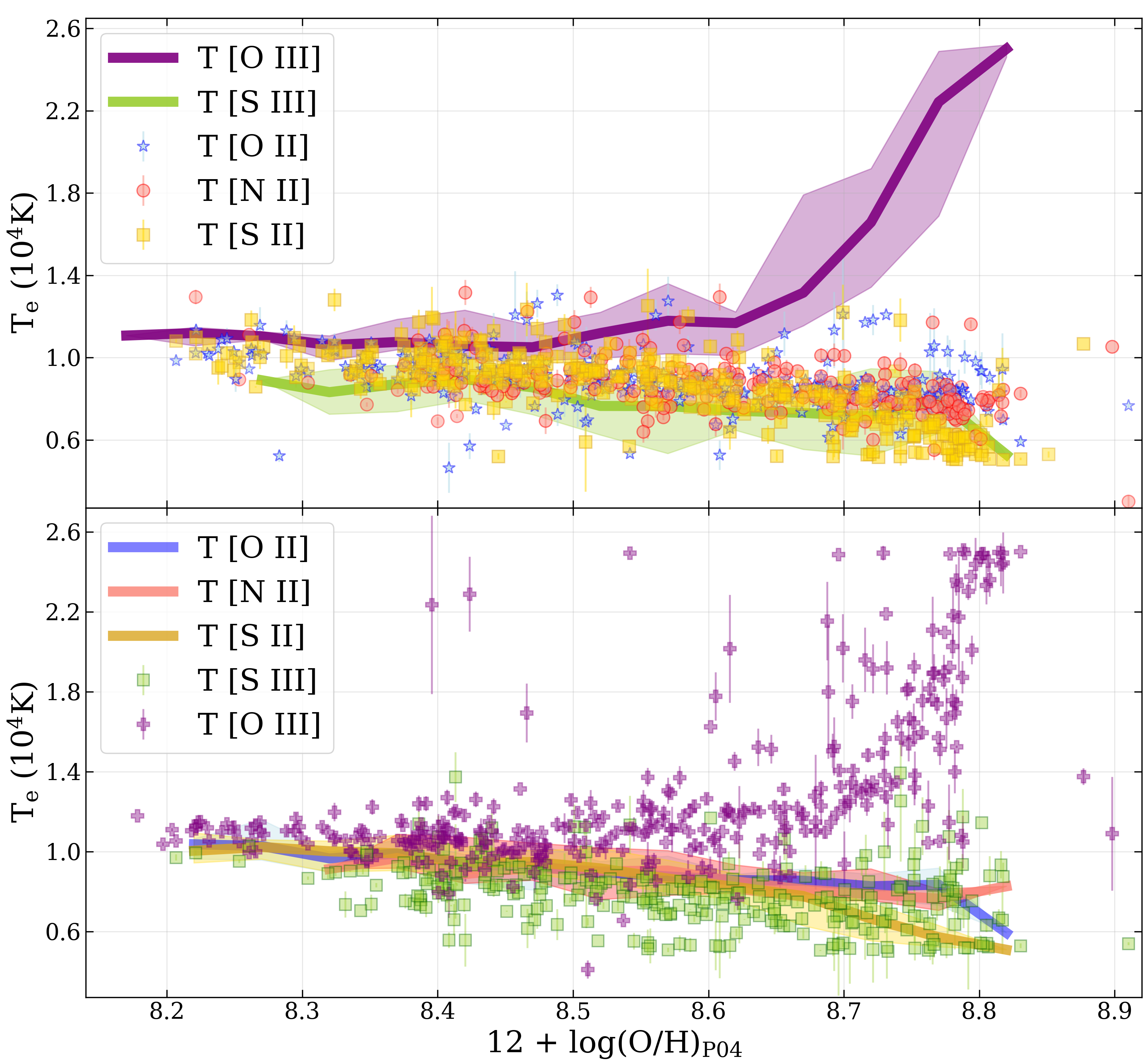}
    \caption{The ionic electron temperature across all SFR--$\mathrm{M_\star}$ and radial bins as a function of strong-line metallicity (P04). Both panels present identical temperature estimates. In the upper panel, OIII and SIII temperatures are binned in metallicity intervals of 0.05 dex, where filled regions represent the 1$\sigma$ percentile of distributions around the median values indicated by solid lines. The same binning procedure is applied to the low-ionization ionic temperatures in the lower panel. The error bars for OII, SII, and NII temperatures in the upper panel as well as OIII and SIII in the lower one correspond to the formal error obtained through the bootstrapping analysis, as detailed in sec.\ref{temp and density section}.}
    \label{fig:ion_temp}
\end{figure*}

To measure electron densities we use the PyNeb software package \citep{2012IAUS..283..422L, 2015A&A...573A..42L} and the [S II]$\lambda$6717,6731 doublet ratio. The majority of our stacks have low electron densities (\(n_e \leq 100 \mathrm{cm}^{-3}\)) since the detected [S II] ratios are usually close to the theoretical limit of 1.41 with $\sigma \sim 0.12$. Therefore, we set \(n_e\) to $100 \mathrm{cm}^{-3}$ for all bins, which is the typical electron density in HII regions \citep{1989agna.book.....O}. The use of a lower density value does not affect the inferred temperatures.

We infer electron temperatures using PyNeb, and temperature-sensitive auroral-to-nebular ratios. To determine the temperatures of low-ionized atoms, we use the [OII]$\lambda\lambda 7318,7319,7329,7330/$[OII]$\lambda\lambda 3726,3728$ ratio for [OII], the [SII]$\lambda\lambda 4068,4076/$[SII]$\lambda\lambda 6716,6730$ ratio for [SII], and the [NII]$\lambda 5754/$[NII]$\lambda 6583$ ratio for [NII]. We use the [OIII]$\lambda 4363/$[OIII]$\lambda 5006$ ratio to obtain [OIII] temperatures, which traces the high-ionization zone,  and the [SIII]$\lambda 6312/$[SIII]$\lambda\lambda 9068,9530$ ratio for [SIII] temperature, which traces the intermediate-ionization zone. To derive uncertainties we add random Gaussian noise consistent with the flux uncertainties and run 1000 Monte-Carlo realisations to obtain the best-fit temperature and its uncertainty, the latter taken as standard deviation of the distribution. The temperature with uncertainties below $500$K are considered reliable for determining ionic abundances. 

To compare with the abundances obtained via the `direct' method we also compute strong-line metallicity estimates for each bins based on strong nebular lines. We used as a fiducial estimate the one obtained from  \cite{2004MNRAS.348L..59P}, P04 hereafter, based on the O3N2 index, defined as O3N2 $\equiv$ $\log( (\Oiii/H\beta)/(\Nii/H\alpha ) )$. The O3N2 index offers the advantage of reducing uncertainties stemming from flux calibration or extinction correction issues, as it relies on line ratios with similar wavelengths that vary monotonically with metallicity. The uncertainties in the strong-line estimates of metallicity are also estimated via Monte-Carlo simulations.

In Fig. \ref{fig:ion_temp}, the temperatures of different ions are shown as a function of P04 metallicity for 336 (56$\times$6) bins. Both panels present the same dataset. In the upper panel, OIII and SIII temperatures are grouped into metallicity bins of 0.05 dex. The lower panel follows the same binning procedure for low-ionization ionic temperatures. It is evident that ionic temperatures generally decrease with increasing metallicity, starting from approximately $11,000\mathrm{K}$ at the lowest metallicities and dropping to around $6000\mathrm{K}$ at $\log(\mathrm{O/H})$ > 8.75, except for [OIII]. 

The behaviour of the [OIII] temperature in Fig. \ref{fig:ion_temp} is noticeably divergent from that of the other temperatures. At low metallicities, the [OIII] temperatures are consistently higher than that of the low-ionization lines (by $600\mathrm{K}$ on average). At higher metallicity, however, the inferred [OIII] temperature increases to very high value, with an inflection point around $12+\log(\mathrm{O/H})= 8.5$ as inferred from the P04 calibration. We consider this effect as potentially associated with imperfections in the flux measurement in the presence of strong contamination at high metallicity, as discussed above. Alternatively, it could be associated with shocks or additional physical processes besided photoionization which may primarily affect the innermost, most highly ionzied, section of the nebula \cite{RickardsVaught2024}.
Whatever the origin of this behaviour, its effect on the inferred [OIII] temperatures is significant enough to make us question their reliability. In this work we do not employ [OIII] temperatures in ionic abundance measurements, deeming them unreliable.
A more accurate measurement of the [OIII] temperature requires either high spectral resolution observations, capable of identifying the sources of contamination in the high-metallicity regime, or a more accurate model for the physical source of the contamination emission.

\begin{figure*}
    \centering
    \includegraphics[width=\linewidth,height=0.905\textheight,keepaspectratio]{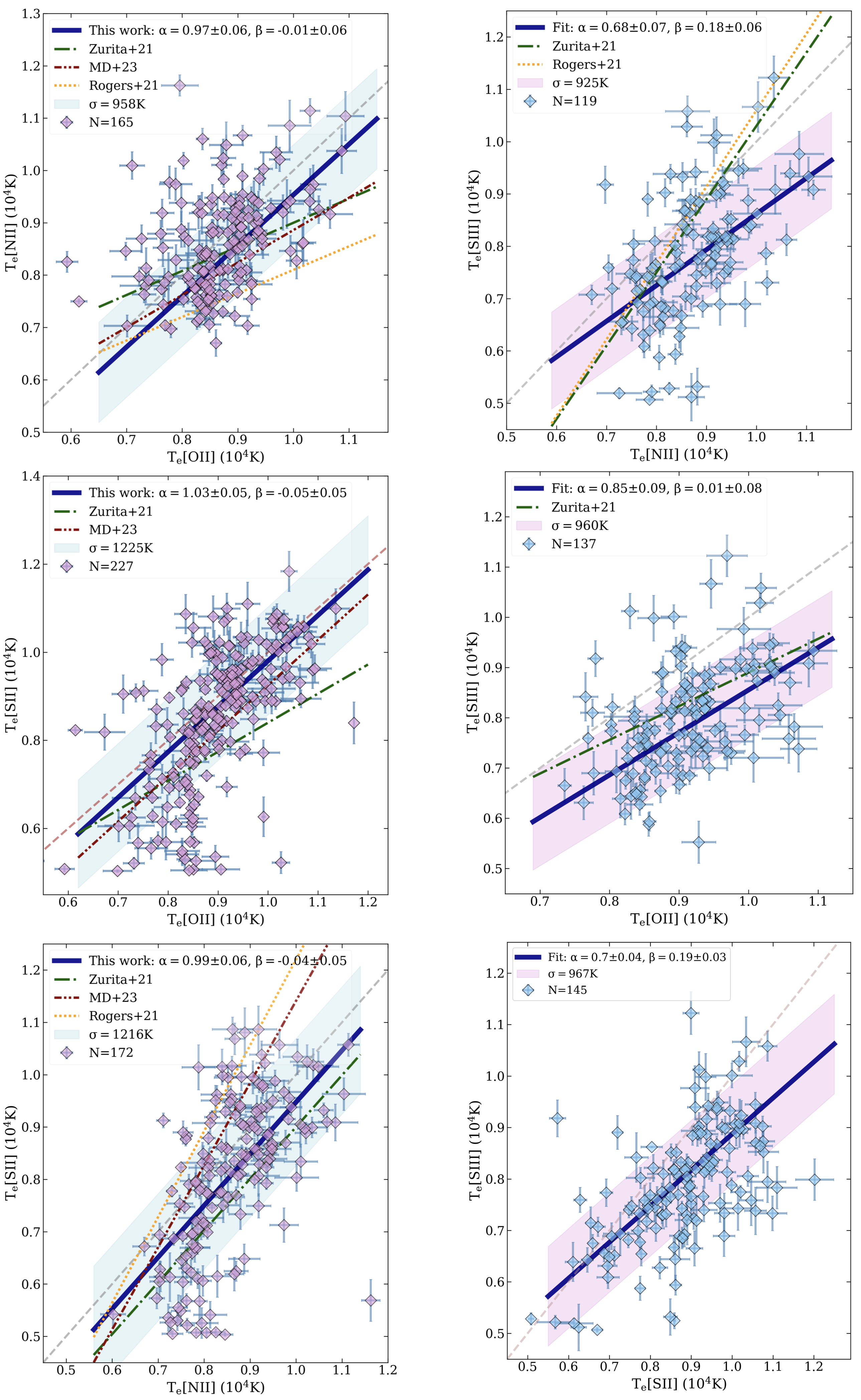}
    \caption{The T$_e$ data for [O II], [N II], [S II], and [S III] auroral lines for our galaxy sample in 56 SFR--$\mathrm{M_\star}$ bins in 6 different radii. The columns are ordered by ionization zone relations: the left panels, low-ionized zones T$_e$ are compared, and the right panels, low- and high-ionized zones T$_e$ are compared. The solid lines indicate the linear fit to error-weighted sigma-clipped temperatures, and the dashed lines represent the one-to-one relation. The shaded regions ($\sigma$) represent the standard deviation of temperatures distributed around the fitted values. Each plot specifies the number of data points, and error bars represent the official temperature errors outlined in sec.\ref{TT relations}. The green and red dashed-dotted lines represent the T-T relations introduced by \cite{2021Zurita} and \citeauthor{2023Mendez-Delgado} (MD+23, \citeyear{2023Mendez-Delgado}), respectively, while the orange dotted line shows the relation reported by \cite{2021Rogers}.
    }
    \label{fig:TT_LowIonized}
\end{figure*}

\subsection{T-T Relations}\label{TT relations}

The temperature of various ionization zones in HII regions is probed by different ions \citep[e.g.,][]{Garrnet1992,2003Kennicutt}. In this section we explore the relations between different electron temperatures (T-T relations) in 336 (56$\times$6) bins. We adopt a  bootstrapping analysis to mitigate the potential effects of the stacking process on our temperature determinations. In particular, for each radial bin in the SFR--$\mathrm{M_\star}$ plane we randomly select 90\% of the galaxies in the bin and generate 150 different bootstrapped stacks. We measure the line fluxes in each bootstrapped realization and use these fluxes to infer temperatures. Only realizations with temperature uncertainties below $200$K are considered reliable. Subsequently, the standard deviation obtained from the bootstrapped reliable temperatures serves as the formal error estimate for each bin. Notably, these formal errors are several orders of magnitude larger than the actual temperature errors obtained from flux uncertainties.

As a result, we measure the electron temperatures encompassing a broad range of temperatures from $\sim 5000 \mathrm{K}$ to approximately $12,000 \mathrm{K}$. In fitting the T-T relations we consider only temperatures with formal errors $<500\mathrm{K}$ and weight the data points by their uncertainties in both axes using inverse variance weighting. Additionally, we implement a sigma clipping method ($2\sigma$) on the error-weighted temperatures to reduce the impact of outliers on the determination of the best fits.

For low-ionisation ions ([OII], [NII], [SII]) we see a strong correlation and fairly close to one-to-one relation between their respective temperatures (Fig. \ref{fig:TT_LowIonized}). The resulting best fits are consistent with linear relations with zero or small intercepts: 

\begin{equation}\label{eq 1 O2N2 tt}
    \mathrm{T_{e}[NII]} = (0.97\pm0.06) \times \mathrm{T_{e}[OII]} + (100\pm600) K;
\end{equation}
\begin{equation}\label{eq 2 S2O2 tt}
    \mathrm{T_{e}[SII]} = (1.03\pm0.05)\times \mathrm{T_{e}[OII]} - (500\pm500) K;
\end{equation}
\begin{equation}\label{eq 3 S2N2 tt}
    \mathrm{T_{e}[SII]} = (1.04\pm0.06)\times \mathrm{T_{e}[NII]}- (900\pm600) K.
\end{equation}

The quadruplet of auroral lines [O II]$\lambda\lambda\lambda\lambda7318,7316,7329,7330$ is located in a region of the optical spectrum which suffers from extensive contamination from sky lines. The measurements of their flux may also be affected by phenomena including collisional de-excitation, imperfect reddening correction, and telluric emission from OH bands \citep[e.g., ][]{2003ApJ...591..801K,2006MNRAS.370.1928P}. These factors, in turn, introduce potential uncertainties the [OII] temperature, which has been considered in the literature as the least reliable among the low-ionization lines. In this work, however, we find excellent agreement between the low-ionization line temperatures, which models expect to be identical in a wide range of ionized gas conditions (e.g., \citealt{2013ApJ...765..140A,2016ApJ...830....4C})

The [SIII] ion probes the intermediate ionization zone and its temperature is compared with that of the low-ionization ions in the right panels in Figure \ref{fig:TT_LowIonized}. We find a moderate to strong positive correlation between \( \mathrm{T_e[S  III]} \) and the $T_e$ of low-ionization ions, with an average correlation coefficient of 0.53. We also observe a consistent deviation from the one-to-one relation, with [SIII] generally showing lower temperatures than [OII] and [SII]. The deviation between the [SIII] and low-ionization temperatures is larger at high temperatures, i.e. lower metallicities. Our findings indicate that \( \mathrm{T_e[O  II]} \) exhibits a more rapid increase compared to \( \mathrm{T_e,[S  III] }\), align more closely with the empirical trend identified by \cite{2021Zurita}, while, according to the predictions of the \cite{2016ValeAsari} photoionization models, the temperature \( \mathrm{T_e[S  III]} \) is expected to be slightly higher than \( \mathrm{T_e[O  II]} \) by a consistent margin throughout the entire range of temperatures.

These relations have also been empirically established in previous investigations \citep[e.g.,][]{2009ApJ...700..654E,2009MNRAS.398..485P,2013ApJ...765..140A,2016ApJ...830....4C,2021Zurita,2020A&A...634A.107Y,2020Berg,2021Rogers,2024Rickards-Vaught}. The results presented in this section are generally in qualitative agreement with previous studies. Specifically, our findings align with those of \cite{2020Berg,2021Zurita} for both low and high ionization zones, and with \cite{2021Rogers,2023Mendez-Delgado} for low ionization zone temperatures. Additionally, the results of \cite{2024Rickards-Vaught} are compatible with our findings regarding the comparison between [SIII] and low-ionization ionic temperatures.

\subsection{Ionic Abundances Calculations}\label{ionic abundance calc}

We calculate ionic abundances using Pyneb, using electron temperature, electron density, and the appropriate flux ratio of the nebular emission line with respect to H$\beta$.

To calculate the abundance of O$^{+}$ we use the [O II]$\lambda\lambda$3726,3729 nebular lines and the error-weighted average temperature of low-ionized ions ([NII], [OII], and [SII]). This approach is supported by our finding in Sec. \ref{TT relations} that
\begin{equation}\label{Equi_Temps}
 T_{\rm [OII]} \approx T_{\rm [SII]} \approx  T_{\rm [NII]}.  
\end{equation}

Furthermore, despite the high signal-to-noise ratio (S/N) of our stacked spectra, some stacks have OII or other low-ionization ion temperatures with errors exceeding the cutoff of $500\mathrm{K}$. The averaging approach allows us to fold in this relatively weak detections by using alternative temperature measurements when one, especially OII, is excluded.

\begin{figure}
    \centering
    \includegraphics[width=\linewidth,height=\textheight,keepaspectratio]{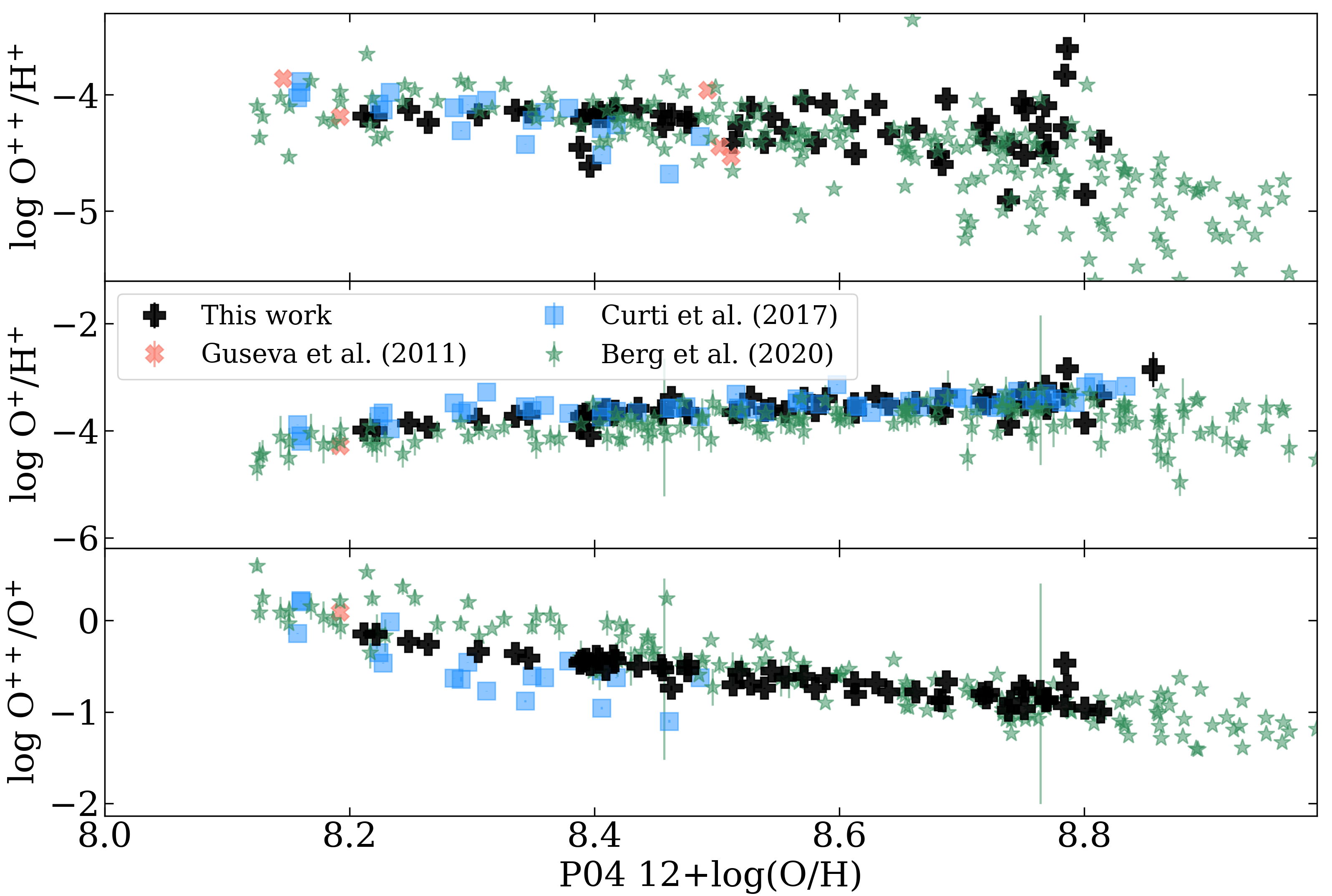}
    \caption{The O$^{++}$ and O$^{+}$ ionic abundances are demonstrated as a function of metallicity measured from the \cite{2004MNRAS.348L..59P} O3N2 calibration. Our measurements of MaNGA data are compared to ionic abundances that are reproduced by employing emission line fluxes that are reported in \cite{2011Guseva}, \cite{2017MNRAS.465.1384C}, and \cite{2020Berg}. Note, ionic abundances, global and radially binned, are available online as introduced in Tab.\ref{tab:Radially binned inferred} and \ref{tab:Global inferred physical}. }
    \label{fig:Ion_Comparison}
\end{figure}

To estimate O$^{++}$ abundances we use the [O III]4959,5007\AA\, fluxes and infer the temperatures of [OIII] from that of the other ions applying the T-T relations from  \cite{2021Rogers}. In particular, we obtain our best estimate of the OIII temperature by using the median the temperatures obtained using the T-T relations from \cite{2021Rogers} relating T[OIII] with T[SIII] 
\begin{equation}
    T_{\rm [O III]} = 0.63 \times T_{\rm [S III]} + 3600,
\end{equation}
and with the temperature of each of the measured low-ionisation ions
\begin{equation}
    T_{\rm [O III]} =  1.30 \times T_{\rm low \ ion} - 2000.
\end{equation}

We compute the ionic abundances of O$^{+}$ and O$^{++}$ ions for each stacked spectrum. Finally we compute the total oxygen abundance as the sum of these two ionic abundances neglecting higher ionization states. Moreover, through the use of MC simulations, we calculate the inferred oxygen abundance uncertainties by randomly perturbing all recorded line fluxes 1000 times, based on the assumption of a Gaussian noise distribution and their measured errors. We only consider temperatures with uncertainties below 500 K to categorize bins with well-detected auroral lines.

In Fig.\ref{fig:Ion_Comparison}, show the trends of O$^{+}$ and O$^{++}$ abundances (top panels) and their abundance ratios (bottom panel) with strong-line metallicity. 
As expected, O$^{++}$/H$^{+}$ ratio decreases with metallicity, while the O$^{+}$/H$^{+}$ ratio shows a positive slope in the whole metallicity range. The bottom panel of Fig. \ref{ionic abundance calc} demonstrates a decreasing O$^{++}$ to O$^{+}$ ratio with increasing metallicities. Incidentally, because of the lower O$^{++}$ abundance at high metallicity, the overall oxygen abundance measurements are not very sensitive to the choice of T[OIII] in the high-metallicity regime.

We compare these abundance trends with those obtained using data from the literature extracted from \cite{2011Guseva}, \cite{2017MNRAS.465.1384C}, and \cite{2020Berg}. \cite{2011Guseva} used archival VLT/FORS1+UVES spectroscopic data to study a large sample of star-forming galaxies with low metallicity, encompassing a total of 121 spectra. The \cite{2017MNRAS.465.1384C} data consists of stacks of SDSS galaxies, while \cite{2020Berg} examined 190 individual HII regions in  NGC~628, NGC~5194, NGC~5457, and NGC~3184, which were targeted by Multi-Object Double Spectrographs \citep[MODS,][]{2010Pogge} on the Large Binocular Telescope (LBT) as a part of the on-going survey of the CHemical Abundances of Spirals \citep[CHAOS, ][]{2015Berg}. We use the emission line fluxes reported in these literature studies but recomputed chemical abundances following the same procedure adopted in this work. Our results are consistent with O$^{+}$ and O$^{++}$ abundances derived from those studies. Furthermore, we confirm that our findings remain consistent with these studies even when using the [OII] and [OIII] temperatures derived from their reported fluxes.

\section{RESULTS}\label{sec result}

\subsection{Mass-Metallicity Relation}

\begin{figure}
    \centering
    \includegraphics[width=\linewidth]{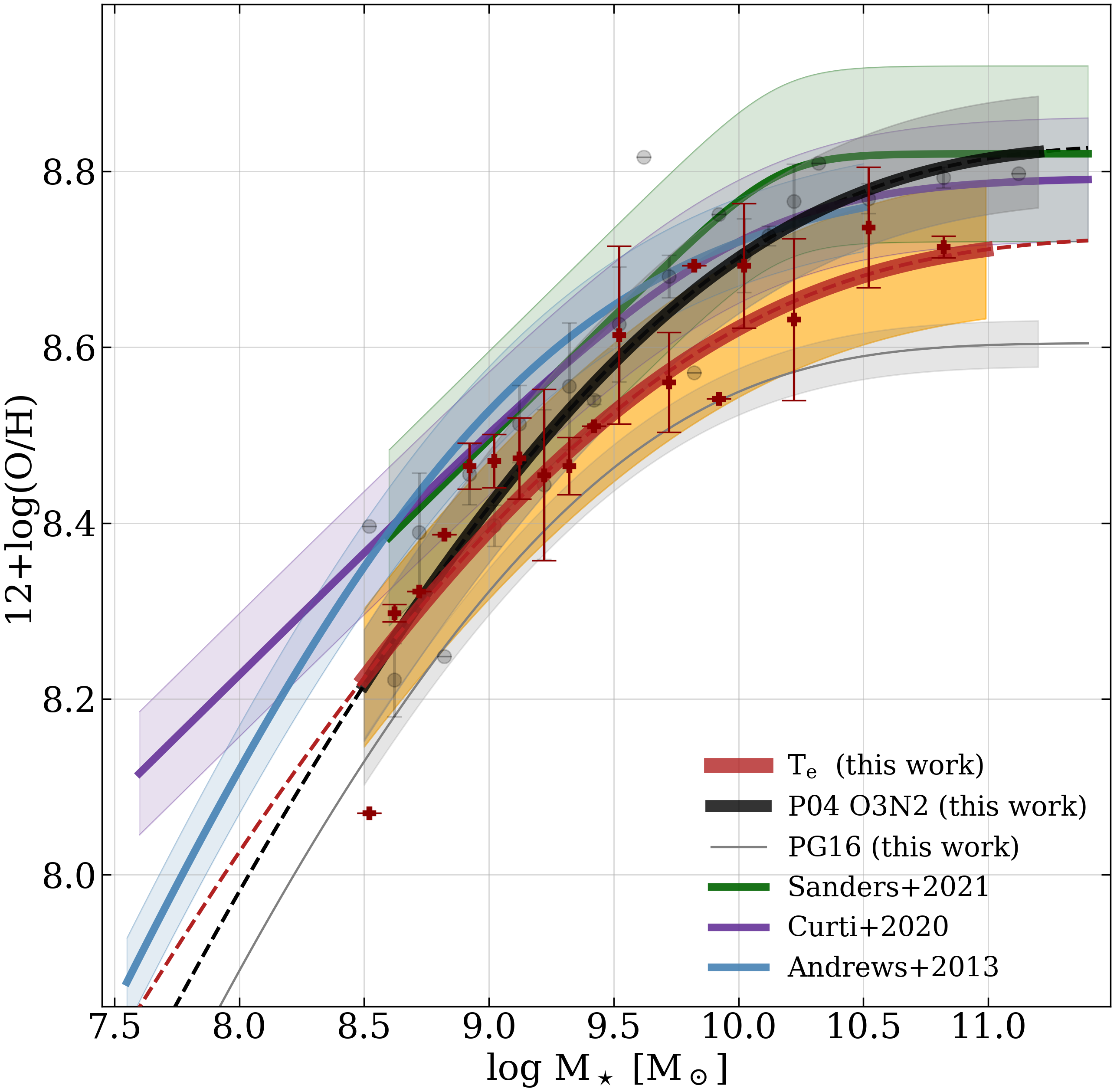}
    \caption{ The mass-metallicity relation obtained using $T_e$-based metallicity from the stacked spectra (brown solid line) compared to the MZR obtained using strong-line methods (P04 O3N2 and PG16) and relations from the literature. Direct method measurements (8.5$< \text{log M}_\star/ M_\odot<11.0$) are shown in in brown, with the line representing the best fit following the equation \ref{MZR_eq} functional form, and the shaded region indicating the standard deviation of metallicity distributed around the fit. Metallicity measurements via empirical calibrations (PG16 and P04) from our sample are represented by gray and black colors, respectively. The comparisons with the characterizations of the MZR from \cite{2013ApJ...765..140A}, \cite{2020Curti}, and \cite{2021Sanders} are depicted by the blue, purple, and green lines, respectively, along with their corresponding standard deviations. Note, metallicity values, global and radially binned, are available online as introduced in Tab.\ref{tab:Radially binned inferred} and \ref{tab:Global inferred physical}.}
    \label{fig:MZR}
\end{figure}

As shown in Fig.\ref{fig:MZR}, we present the mass-metallicity relation obtained from our estimates of direct metallicity using SFR--$\mathrm{M_\star}$ bins and stacking all the spectra in the 0.0-1.5$R_e$ radial range. This leads to 54 measurements of metallicity as a function of M$_\star$ and SFR. The shaded areas represent the standard deviation of the entire sample (individual SFR--$\mathrm{M_\star}$ bins) from the fit. For ease of visualization in \ref{fig:MZR}, we divide the sample into mass bins of 0.1 dex width, with error bars representing the standard deviation of the metallicity distribution. In Fig. \ref{fig:MZR}, we present a fit to metallicity values, adopting a slightly modified version of the \cite{2014Zahid} functional form, as introduced in \cite{2020Curti}, following the equation:
\begin{equation}
    12 + \mathrm{log(O/H)} = \mathrm{Z}_0 + \mathrm{log}(1-10^{-(\frac{\mathrm{M_\star}}{\mathrm{M}_0})^\gamma}).
    \label{MZR_eq}
\end{equation}

Our best-fitting median MZR, determined using the direct method, asymptotes at $12 + \mathrm{log(O/H)} = 8.73 \pm 0.04$, with a turnover at $\mathrm{log(M_\star/M_\odot) = 9.99 \pm 0.18}$ and a low-mass-end slope of $\gamma = 0.26 \pm  0.07$, as reported in Table \ref{tab:MZR_fit}. We also show the results of the same analysis on our sample using the P04 O3N2 and the PG16 \citep{2016MNRAS.457.3678P} strong-line metallicity calibration.

\begin{table}
\caption{Best-fitting values for the parameters of the MZR, according to Eq. \ref{MZR_eq}, derived with the direct method, and the calibrations of P04 and PG16. }
\resizebox{\columnwidth}{!}{%

\begin{tabular}{rllll}

$12+\mathrm{log(O/H)}$&$\mathrm{Z}_0$&$\mathrm{log(M_0/M_\odot)}$&$\gamma$\\
\hline
$\mathrm{T}_e$&
$8.73 \pm 0.04 $&
$9.99 \pm 0.18 $&
$0.26 \pm  0.07$\\

P04 &
$8.86 \pm 0.04$&
$10.13 \pm  0.23$&
$0.29 \pm 0.03$\\

PG16 &
$8.60 \pm 0.02$&
$9.68 \pm  0.09$&
$0.32 \pm 0.03$\\

\end{tabular}
}

\label{tab:MZR_fit}
\end{table}

In previous studies, it has been shown that the shape of the local MZR exhibits notable differences depending on the metallicity indicator and calibration used \citep{2008Kewley,2012López-Sánchez,2020Curti}. Therefore, the comparison of metallicities derived from different indicators and calibrations has the potential to introduce substantial biases. Nevertheless, our directly measured MZR, with a standard deviation of 0.08 dex, is both qualitatively and quantitatively consistent with the calibration values of P04 (0.07 dex standard deviation) within the associated uncertainties. The shape of the $T_e$-MZR also agrees with PG16, albeit the absolute values provided by PG16 are consistently lower than other measurements, with metallicities barely reaching a maximum of 8.65. This behavior has been pointed out in previous studies, adopting a MaNGA sample \citep[e.g.,][]{2022Puertas,2023Boardman}.

\begin{figure*}
    \centering
    \includegraphics[width=\linewidth ,height=0.95\textheight,keepaspectratio]{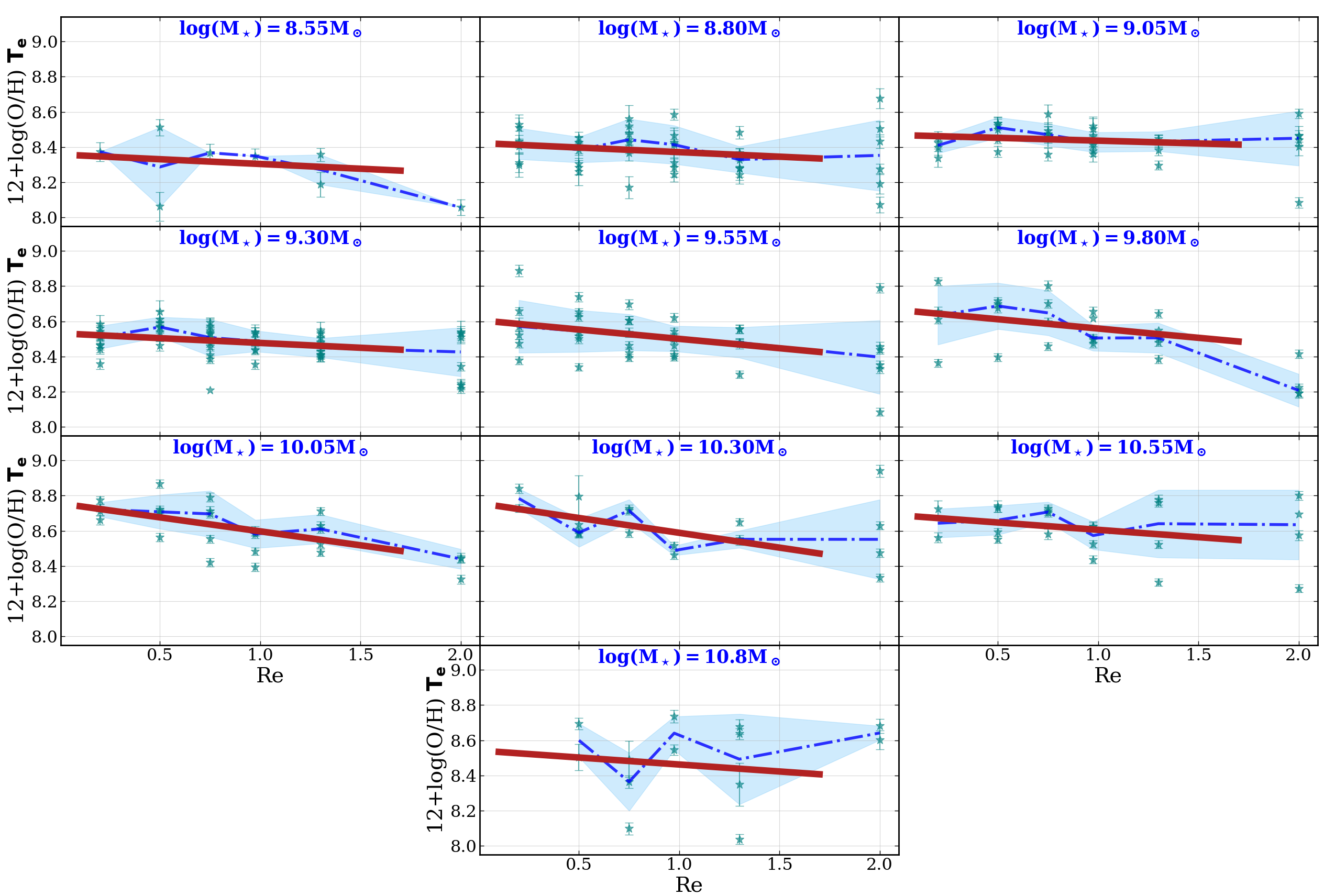}
    \caption{The metallicity gradient of  SFR--$\mathrm{M_\star}$ bins with different stellar masses, separated by 0.25dex. In each panel, the solid line represents a linear fit to SFR--$\mathrm{M_\star}$ bins far out to 1.5Re. The dash-dotted line demonstrates the trend of the median metallicity of all bins up to 2.5$R_e$ and The shaded region shows the standard deviation of metallicity distribution around the dash-dotted line. }
    \label{fig:Amet Grads}
\end{figure*}

Fig.\ref{fig:MZR} also shows a comparison with characterizations of the MZR from the literature \citep{2013ApJ...765..140A,2020Curti,2021Sanders}. Using stacked spectra from SDSS \cite{2013ApJ...765..140A} directly measure oxygen abundances for SFR--$\mathrm{M_\star}$ bin. The MZR by \cite{2020Curti} encompasses a mass range of $10^{8}-10^{11.5}$ M$_\odot$ and is also based on the $T_e$-derived metallicities from SDSS stacked spectra, albeit binned in the [OIII]$\lambda$5007/H$\beta$ versus [OII]$\lambda$3727/H$\beta$ plane. Lastly, \cite{2021Sanders} expanded upon the galaxy sample used by \cite{2013ApJ...765..140A}, incorporating the sample of 38 dwarf galaxies from the Spitzer Local Volume Legacy survey \citep{2020Berg} within a mass range of $10^{8.5}-10^{11.5}$ M$_\odot$. They took into consideration the contributions from diffuse ionized gas (DIG) to the total emission line fluxes in integrated galaxy spectra, a factor they demonstrated to lead to an overestimation of metallicity. This comparison shows that our directly measured metallicities using the MaNGA sample are qualitatively consistent with those from other studies, which employ different samples and various metallicity diagnostics.

\subsection{Metallicity Gradients}
\label{met_grads}

Fig.\ref{fig:Amet Grads} shows the $T_e$-derived metallicity profiles out to 2.5 $R_e$ in 10 bins of $M_\star$ from $\sim$10$^{8.4}$ to $\sim$10$^{11}$ M$_\odot$. Note that, To reduce scatter and ensure smaller bin ranges at higher masses, Fig.\ref{fig:Amet Grads} replaces the original mass bins from Fig.\ref{fig:bins} with 0.25 dex bins, using the median mass of galaxies in each $\mathrm{SFR-M_\star}$ bin. The dash-dotted lines show the median metallicity in each radial bin while the shaded regions show the standard deviation of data points of different SFR around the median. We can measure $T_e$ metallicity even in the farthest radial bin, 1.5-2.5 $R_e$. 
However, to obtain more robust metallicity gradient slopes, we fit only radial bins up to 1.5 $R_e$. This choice aligns with the common radius covered by MaNGA across the entire sample (see sec.\ref{sec Data}). In Fig.\ref{fig:Amet Grads} the solid lines represent the best fit linear relation between oxygen abundance and radius normalized to $R_e$. In most cases the extrapolation of the linear fit agrees well with the measurements in the 1.5-2.5 $R_e$ radial bin.

\begin{figure*}
    \centering
    \includegraphics[width=\linewidth]{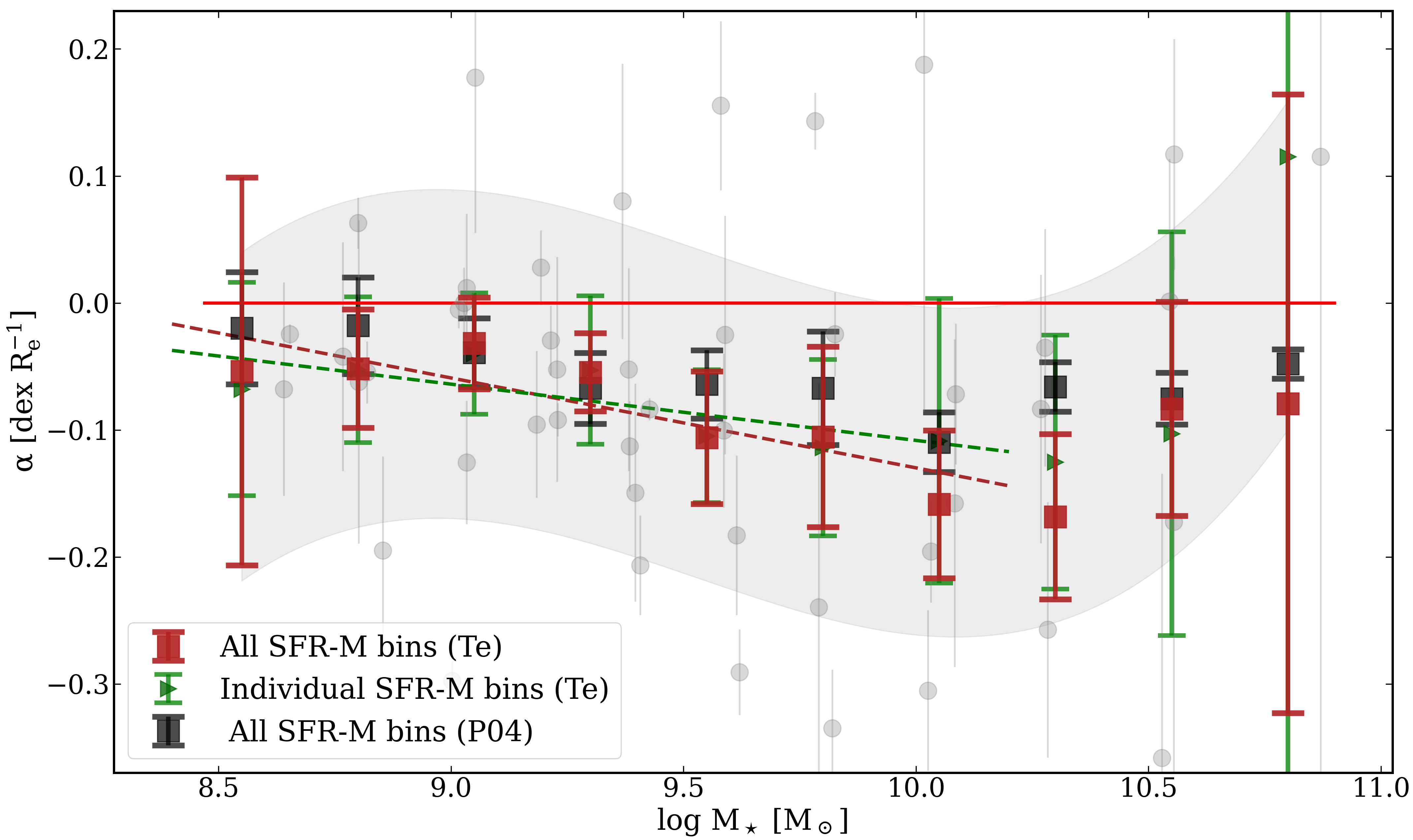}
    \caption{ The metallicity gradient across a radial span of 0.0 to 1.5$R_e$ using the Te (red) and P04 calibration based on O3N2 (black) as a function of stellar mass. The red data points correspond to the slopes depicted in Figure \ref{fig:Amet Grads}, with error bars indicating the uncertainty in the slope. Each $\mathrm{SFR-M_\star}$ bin's slopes are shown with gray points, while green points represent the median value of slopes within a specific mass range. To visually clarify, the gray shaded area represents a third-degree polynomial fit to the slopes of all SFR-M bins.}
    \label{fig:Gradient slopes}
\end{figure*}

Fig. \ref{fig:Gradient slopes} illustrates the correlation between the metallicity gradient and $M_\star$, MZGR. In this figure, we show the values derived from the linear fits (red) in Fig. \ref{fig:Amet Grads} and the median slope of individual SFR--$\mathrm{M_\star}$ bins within the specified mass range (green). We observe that, in both approaches, the slope of metallicity gradients exhibits variations across different stellar masses. It transitions from being relatively flat or slightly positive at $\log(M_\star/M_\odot) \lesssim 9.2$ to negative, reaching the most negative value around  $M_\star \sim 10^{10.3} M_\odot$. In bins with higher mass, there is a reversal of this trend, with gradients appearing shallower (see Tab.\ref{tab:slopes}). This observation is further validated by performing a third-order fit to the slopes of individual SFR--$\mathrm{M_\star}$ bins, with the standard deviation of these slopes illustrated by the gray-shaded region. 

Similar qualitative characteristics in the shape of the metallicity profile are identified when utilizing the R23\footnote{R23 = ([O II]$\lambda$3726,28+[O III]$\lambda$4959,5007)/H$\beta$} metallicity calibrators of \cite{2008Maiolino} and \cite{2021Sanders}.

\section{Discussion}\label{discussion}

\subsection{Metallicity Gradient}

The spatial resolution of MaNGA (1–2 kpc) can introduce a flattening of measured metallicity gradients due to resolution effects \citep{2013Yuan,2018Carton,2020Acharyya}. However, this resolution remains largely consistent across the sample, varying minimally with stellar mass except at the highest mass end ($\mathrm{\log M_*/M_{\odot} > 10.5}$). Consequently, comparisons of metallicity gradients across different mass bins are expected to yield reliable differential measurements, with any resolution-induced bias affecting all but the most massive galaxies similarly.

Our findings are aligned with the \cite{2017BelfioreO&N} study where they found non-linear dependence of the metallicity gradient on stellar mass ($10^9-10^{11}$ M$_\odot$) with a smaller MaNGA (DR13) sample size and employing strong-line calibrations. Moreover, using SAMI survey data, \cite{2018Poetrodjojo} corroborated the variations in the metallicity gradient with increasing stellar mass within a range of $10^9-10^{10.5}$ M$_\odot$. Our results are also consistent with the findings of \cite{2020Mingozzi} using MaNGA data. They demonstrated a strong mass dependency of metallicity gradients, with galaxies in the mass range $10.2 \lesssim \mathrm{log(M_\star/M_\odot)} \lesssim 10.4$ exhibiting the steepest negative gradients, employing various metallicity calibrations.

In this work, we aim to avoid oversimplifying the physical mechanisms underlying metallicity gradients and MZGR behavior, as a complete understanding requires both observational and theoretical perspectives. For instance, regarding the observed flattening of the metallicity gradient at the low-mass end, some studies attribute this effect to stellar feedback, which regulates metallicity before significant ISM and metal mixing occurs \citep[e.g.,][]{2018Chisholm,2021Sharda}. Alternatively, other studies suggest that gas mixing and wind recycling in the outer regions of galaxies may drive this flattening \citep[e.g.,][]{2017BelfioreO&N}.

The region with the steepest gradient and highest curvature around $\mathrm{10^{10-10.5} M_\odot }$ (shown in Fig.\ref{fig:Gradient slopes}) could provide valuable insights into galaxy evolution and the key factors shaping galaxy characteristics from local to high redshift. For instance, \citet{2021Sharda} demonstrated that galaxies dominated by advection tend to flatten at the low-mass end, whereas accretion-dominated galaxies exhibit flattening at the high-mass end. Notably, they observed that advection and accretion do not simultaneously weaken, thereby ruling out diffusion as a primary gradient-smoothing mechanism. Their model indicates that the characteristic bend in the mass--metallicity gradient relation occurs when two major processes that influence gradient smoothing—accretion and inward gas advection—are at their lowest relative to metal production. Although the stellar mass at which this transition occurs may vary depending on model parameters and metallicity calibrations (see \citealt{2021Poetrodjojo}), their findings demonstrate that the presence of this bend is robust across different model assumptions.

At intermediate redshift, \cite{2018Carton} used MUSE data and found marginal evidence for similar metallicity gradient trends (up to M$_\star \, \sim 10^{10.5}$ M$_\odot$) for a selection of 84 galaxies (0.1 < z < 0.8) employing a forward-modeling technique. At higher redshifts, \cite{2024Cheng} illustrated weak but significant anticorrelation between the metallicity gradient and the stellar
mass in 238 star-forming galaxies at cosmic noon. Furthermore, \cite{2024Venturi} found flat gas-phase metallicity gradients in three systems at $6 < z < 8$, and \cite{2024Vallini}, reported a flat gradient at the epoch of reionization ($\mathrm{z} \sim 7$).

\begin{figure}
    \centering
    \includegraphics[width=\linewidth]{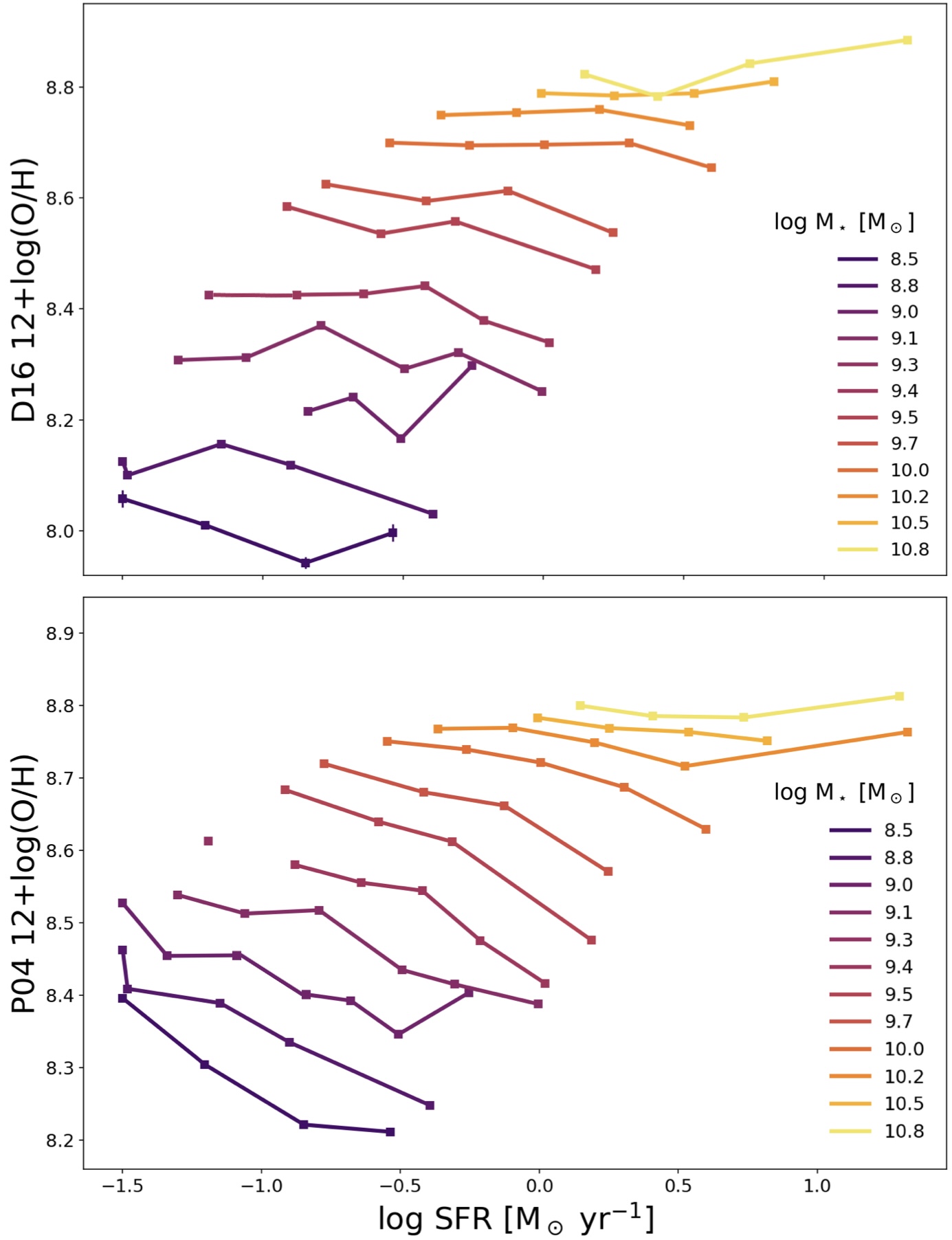}
\caption{The relationship between global $\log(\mathrm{O/H})$ and SFR for our sample is color-coded for different stellar mass bins. In the upper panel, metallicities are calculated using the empirical calibration by \cite{2016Dopita}, while in the lower panel, metallicities are determined via the calibration by \cite{2004MNRAS.348L..59P}. On average, the correlation coefficients between $\log(\mathrm{O/H})$ and SFR are $-0.17$ for the upper panel and $-0.75$ for the lower panel.}
    \label{fig:FMR_P04_D16}
\end{figure} 
\begin{figure}
    \centering
    \includegraphics[width=\linewidth]{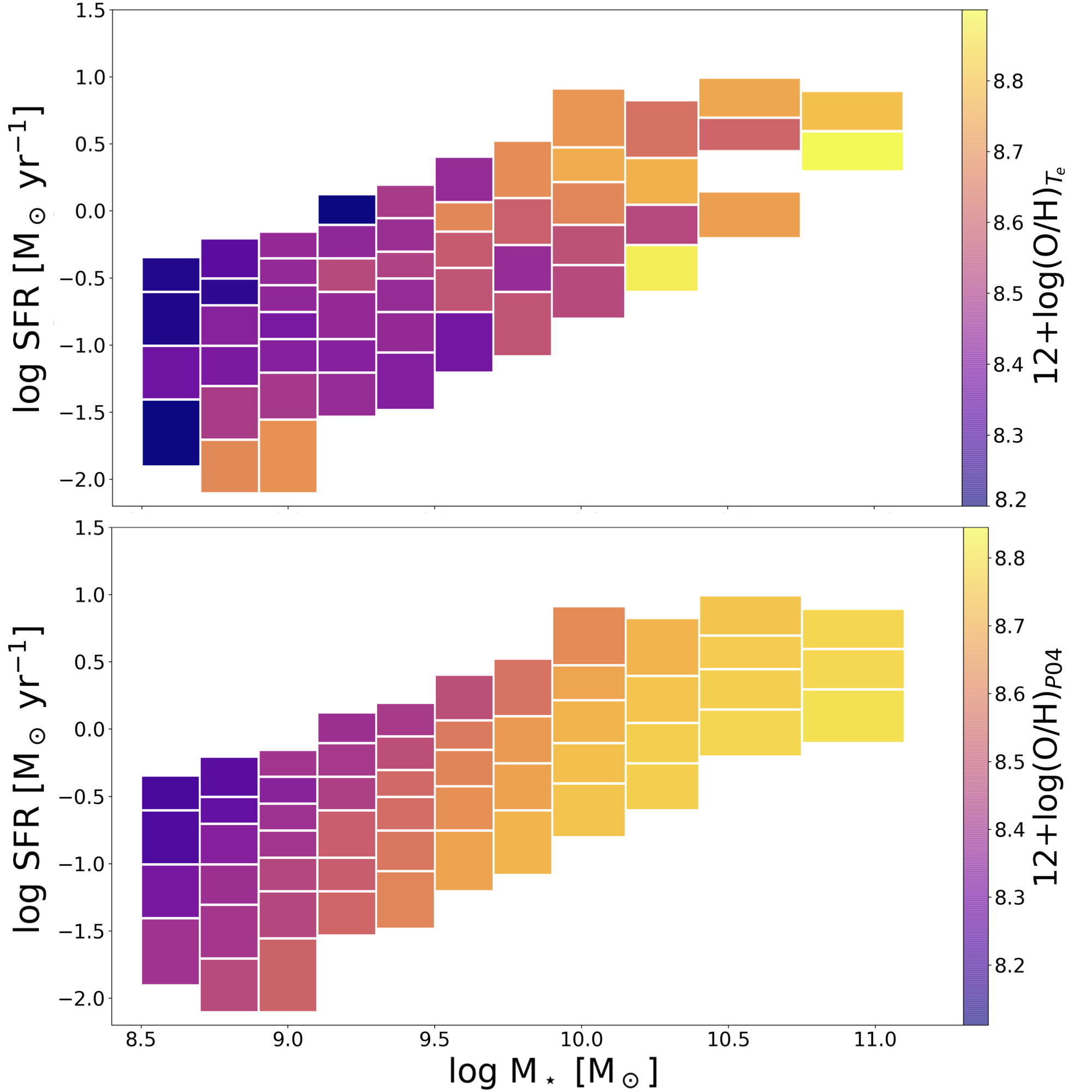}
    \caption{SFR--$\mathrm{M_\star}$ bins are color-coded according to direct metallicity measurements (top panel) and P04 calibrations (bottom panel). Note that any missing bins are a result of either large temperature errors or lower auroral line SNR, which falls below the introduced cutoff (Sect.\ref{sec Data Analysis}). }
    \label{fig:FMR}
\end{figure} 
\subsection{FMR(?)}

The existence of a secondary dependence on SFR in the MZR remains a hotly debated topic in the literature. Several authors have confirmed the existence of the FMR \citep[e.g.,][]{2010Mannucci,2013ApJ...765..140A,2016Hunt,2020Curti,2024Curti,2024Pistis}. Other studies question this conclusion by pointing out that, since most galaxies lie on a tight correlation between $M_\star$ and SFR, the FMR is driven by outliers and including and SFR dependence create a small or negligible reduction in the scatter of the MZR \citep{2017Barrera-Ballesteros, Alvarez-Hurtado2022}.

Moreover, most of these studies are based on subsamples of the same observational data set, the SDSS spectroscopic survey. Despite the application of aperture corrections \citep{2004Brinchmann}, the spectroscopic information is affected by aperture effects \citep[e.g.,][]{2016Gomes,2016Iglesias}. 

Moreover, in a recent study, \cite{2023Baker} explored the dependence of gas metallicity on various galaxy parameters for 1000 star-forming galaxies from the MaNGA survey. Using different statistical methods such as the average dispersion, Partial Correlation Coefficients, and Random Forest regression analysis, they suggest that part of the correlation between SFR and metallicity found in previous studies results from the correlation between gravitational potential and metallicity. 

Recent investigations based on integral field spectroscopy have indicated that local processes, such as those involving gas metallicity, stellar mass surface density, and SFR density, could impact global scaling relations \citep{2016Barrera-Ballesteros,2019Sanchez,2023Baker-M-F}.

Using the P04 O3N2 the existence of the FMR is clearly evident in Figs.\ref{fig:FMR_P04_D16}\&\ref{fig:FMR}, as  metallicity decreases  with increasing SFR across each mass bin. The existence and strength of the FMR depends, however, on the strong-line calibrator used. Using the calibration proposed by \cite{2016Dopita} we obtain a much weaker secondary trend with SFR (see Fig.\ref{fig:FMR_P04_D16}).
On average, the correlation coefficients between $\log(\mathrm{O/H})$ and SFR in fixed mass bins are $-0.75$ for P04 and  $-0.17$  for the calibration of \cite{2016Dopita}.

In Fig. \ref{fig:FMR} we show each bin in the SFR--$\mathrm{M_\star}$ plane colour--coded by its metallicity from the $T_e$ method (top panel) and from the P04 O3N2 strong-line method (bottom panel). 
When employing the direct method (top panel) for measuring metallicity, this correlation largely disappears. In this case the average Pearson correlation coefficient between SFR and metallicity across the different mass bin is $\rho \sim -0.05$.
We tested these conclusion by modifying the bootstrapping procedure, stricter criteria for reliable fluxes and temperatures, using measured [OII] temperatures for determining its abundance, and employing both smaller and larger bins in SFR and M$_\star$ space. However the results are robust to these modifications.

\section{Summary and Conclusion}\label{sec summary and conclusion}

In this study, we have computed the $\mathrm{T_e}$ metallicity across a radial span of 0.0 to 2.5$\mathrm{R_e}$ for 4140 star-forming galaxies, distributed across 56 SFR--$\mathrm{M_\star}$ bins using data from the MaNGA survey (DR17). The enhanced signal-to-noise ratio achieved through the stacking procedure in SFR--$\mathrm{M_\star}$ space enabled us to detect the auroral lines required to determine electron temperatures across various ionization zones, enabling the application of the $T_e$ method to measure direct metallicities.
Here are summarized our main results : 

\begin{enumerate}
  \renewcommand{\labelenumi}{\roman{enumi}.}
  \item  We found excellent agreement between the low-ionization line temperatures traced by [OII], [SII] and [NII], with a nearly one-to-one relationship across a wide temperature range. [SIII], tracing the intermediate ionization zone, generally shows similar or lower temperatures compared to the low-ionization lines. The deviation between [SIII] and the low-ionization temperatures becomes more pronounced at higher temperatures, corresponding to lower metallicities. 

  \vspace{3pt}

  \item The inferred [OIII] temperature are in disagreement with the temperatures measured from other ions, especially at high metallicity. Such deviation is particularly evident  for$12+\log(\mathrm{O/H})> 8.5$ as determined by the P04 calibration. We can potentially attribute this effect to potential imperfections in flux measurement due to significant spectral contamination at high metallicities. Such contamination of the [OIII]4363 line has previously been reported in the literature. While the precise nature of this contamination remains unclear, it is at least partially due to the presence of the [Fe II]$\lambda$4360 line. However, the higher [OIII] temperatures are obtained even after fitting the [Fe II]$\lambda$4360 flux and explicitly constraining to $0.73\times$ that of the nearby isolated [Fe II]$\lambda$4288. A physical origin of the higher T[OIII] temperature (e.g. additional heating in the high-ionization zone) cannot be excluded. In this work we consider the [OIII] temperatures unreliable and exclude them from ionic abundance measurements. 

  \vspace{3pt}
  
  \item We calculated ionic abundances of O$^{+}$ using the low-ionization temperatures and O$^{++}$ based on the other measured electron temperatures.
  The relative abundance of O$^{++}$ to O$^{+}$ decreases with metallicity, indicating a reduced impact of O$^{++}$ or the estimate of T[OIII] on final oxygen abundance estimations at higher metallicities.

  \vspace{3pt}
  
  \item We presented the mass-metallicity relation (MZR) derived from direct metallicity estimates, using SFR-$\mathrm{M}_\star$ bins and stacking spectra in the $0.0-1.5\mathrm{R_e}$ radial range, resulting in 56 metallicity measurements. The measured MZR is in good agreement with the literature.

  \vspace{3pt}
  
  \item We calculated the $\mathrm{T_e}$ metallicity gradient across the radial range of 0.0 to 1.5$\mathrm{R_e}$ for 4140 star-forming galaxies, distributed among 56 SFR--$\mathrm{M_\star}$ bins.  The metallicity gradients revealed a linear relation between oxygen abundances at different radii normalized to $\mathrm{R_e}$, covering stellar masses from approximately $10^{8.4}$ to $10^{11}$ $\mathrm{M_\odot}$. We demonstrated the correlation between metallicity gradient slopes and stellar mass, with slopes transitioning from flat or slightly positive at low stellar masses to negative around $10^{10.3}$ $\mathrm{M_\odot}$, and then becoming shallower at higher masses. This pattern is in agreement with previous studies using smaller MaNGA and SAMI survey samples in the local Universe and with MUSE data at higher redshifts. Our findings are consistent using different metallicity calibrators, including R23 and strong-line methods.

  \vspace{3pt}
  
  \item  We investigated the presence of the FMR. We showed that using strong-line methods we can recover the FMR, although the strength of the FMR signal depends on strong-line calibrator used. However, using the direct method to measure metallicity reveals no clear FMR. 
  
\end{enumerate}

Further spatially resolved investigations may contribute to a better understanding of the FMR and its existence on local scales. Building on this work, stacking spaxels and deriving a relation between gas-phase metallicity, stellar mass surface density, and SFR surface density could address the debated questions regarding the global versus local effects of the FMR, irrespective of the parent galaxy properties. 

\begin{acknowledgements}
AK thanks Amirnezam Amiri, Sirio Belli, Noah Rogers, Erin Kado-Fong, and Mirko Curti for the constructive and insightful discussions that significantly contributed to the development of this work. 

FB acknowledges funding from the INAF Fundamental Astrophysics program 2022.

Funding for the Sloan Digital Sky 
Survey IV has been provided by the 
Alfred P. Sloan Foundation, the U.S. 
Department of Energy Office of 
Science, and the Participating 
Institutions. 

SDSS-IV acknowledges support and 
resources from the Center for High 
Performance Computing  at the 
University of Utah.

SDSS-IV is managed by the 
Astrophysical Research Consortium 
for the Participating Institutions, listed on the SDSS website \url{www.sdss4.org}.

\end{acknowledgements}

\begin{appendix}
\section{Metallicity gradient slopes table}

\begin{table}
\caption{Metallicity gradient slopes as depicted in Fig.\ref{fig:Amet Grads} and Fig.\ref{fig:Gradient slopes}. $\mathrm{T_e}$ (all bins), $\mathrm{T_e}$ (Individual bins), and $\mathrm{P04}$ (all bins) values represent red, green, and black data points in \ref{fig:Gradient slopes}, respectively}

\begin{tabular}{rllllll}

$\mathrm{M_\star}$\-bin &  Method & Slope&  Method & Slope&  Method & Slope\\
\hline
8.55&
$\mathrm{T_e}$ (all bins) &
$-0.054\pm0.150$&
$\mathrm{T_e}$ (Individual bins)&
$-0.068\pm0.084$&
$\mathrm{P04}$ (all bins)&
$-0.020\pm0.044$\\

8.80&
$\mathrm{T_e}$ (all bins) &
$-0.052\pm0.046$&
$\mathrm{T_e}$ (Individual bins)&
$-0.053\pm0.057$&
$\mathrm{P04}$ (all bins)&
$-0.018\pm0.038$\\

9.05&
$\mathrm{T_e}$ (all bins) &
$-0.034\pm0.045$&
$\mathrm{T_e}$ (Individual bins)&
$-0.040\pm0.039$&
$\mathrm{P04}$ (all bins)&
$-0.039\pm0.028$\\

9.30&
$\mathrm{T_e}$ (all bins) &
$-0.055\pm0.031$&
$\mathrm{T_e}$ (Individual bins)&
$-0.053\pm0.058$&
$\mathrm{P04}$ (all bins)&
$-0.067\pm0.028$\\

9.55&
$\mathrm{T_e}$ (all bins) &
$-0.106\pm0.052$&
$\mathrm{T_e}$ (Individual bins)&
$-0.104\pm0.052$&
$\mathrm{P04}$ (all bins)&
$-0.070\pm0.026$\\
  
9.80&
$\mathrm{T_e}$ (all bins) &
$-0.105\pm0.071$&
$\mathrm{T_e}$ (Individual bins)&
$-0.114\pm0.069$&
$\mathrm{P04}$ (all bins)&
$-0.077\pm0.044$\\

10.05&
$\mathrm{T_e}$ (all bins) &
$-0.159\pm0.058$&
$\mathrm{T_e}$ (Individual bins)&
$-0.109\pm0.112$&
$\mathrm{P04}$ (all bins)&
$-0.110\pm0.024$\\

10.30&
$\mathrm{T_e}$ (all bins) &
$-0.168\pm0.065$&
$\mathrm{T_e}$ (Individual bins)&
$-0.125\pm0.099$&
$\mathrm{P04}$ (all bins)&
$-0.067\pm0.020$\\

10.55&
$\mathrm{T_e}$ (all bins) &
$-0.083\pm0.084$&
$\mathrm{T_e}$ (Individual bins)&
$-0.103\pm0.158$&
$\mathrm{P04}$ (all bins)&
$-0.076\pm0.009$\\

10.80&
$\mathrm{T_e}$ (all bins) &
$-0.079\pm0.243$&
$\mathrm{T_e}$ (Individual bins)&
$+0.155\pm0.557$&
$\mathrm{P04}$ (all bins)&
$-0.048\pm0.011$\\
\hline

\end{tabular}

\label{tab:slopes}
\end{table}


\section{Online Catalogs}

\begin{table*}
\caption{Radially binned integrated spectra. }

\begin{tabular}{rll}

Column &  Name & Description\\
\hline
1&
BIN\_ID &
Number of the bin according to fig\ref{fig:bins}, starting  from the lowest $\mathrm{M_\star}$ and SFR\\

2 &
RAD\_BIN&
Indication of the integrated spectra within this radial range, normalized to $\mathrm{R_e}$\\

3 &
Mass&
Median stellar masses estimate of all galaxies within each SFR--$\mathrm{M_\star}$ bin in the unit of $\mathrm{log(M_\star/M_\odot)}$\\

4 &
SFR&
Median star-formation rate of all galaxies within each SFR--$\mathrm{M_\star}$ bin based on MaNGA \textsc{DAP} catalog\\

5 &
Wavelength&
The rest-frame wavelength of bins spectra in the unit of Angstrom\\
  
 6 &
Flux&
Integrated flux within specified radial range in the unit of $\mathrm{10^{-17} erg/s/cm^2/Angstrom}$\\                         
\hline

\end{tabular}

\label{tab:Radially binned integrated spectra}
\end{table*}

\begin{table*}
\caption{Global integrated spectra. }

\begin{tabular}{rll}

Column &  Name & Description\\
\hline
1&
BIN\_ID &
Number of the bin according to fig\ref{fig:bins}, starting  from the lowest $\mathrm{M_\star}$ and SFR\\

2 &
Mass&
Median stellar masses estimate of all galaxies within each SFR--$\mathrm{M_\star}$ bin in the unit of $\mathrm{log(M_\star/M_\odot)}$\\

3 &
SFR&
Median star-formation rate of all galaxies within each SFR--$\mathrm{M_\star}$ bin based on MaNGA \textsc{DAP} catalog\\

4 &
Wavelength&
The rest-frame wavelength of bins spectra in the unit of Angstrom\\
  
5 &
Flux&
Integrated flux within specified radial range in the unit of $\mathrm{10^{-17} erg/s/cm^2/Angstrom}$\\                         
\hline

\end{tabular}

\label{tab:Global integrated spectra.}
\end{table*}

\begin{table*}
\caption{Radially binned emission line fluxes and their uncertainties. The structure and description of the table columns are laid out here, with the complete version available in the online journal.}

\begin{tabular}{rll}

Column &  Name & Description\\
\hline
1&
BIN\_ID &
Number of the bin according to fig\ref{fig:bins}, starting  from the lowest $\mathrm{M_\star}$ and SFR\\

2 &
RAD\_BIN&
Indication of the integrated spectra within this radial range, normalized to $\mathrm{R_e}$\\

3 &
Mass&
Median stellar masses estimate of all galaxies within each SFR--$\mathrm{M_\star}$ bin in the unit of $\mathrm{log(M_\star/M_\odot)}$\\

4 &
SFR&
Median star-formation rate of all galaxies within each SFR--$\mathrm{M_\star}$ bin based on MaNGA \textsc{DAP} catalog\\

5 &
OII3726\_FLUX&
[O II]$\lambda3726\AA$ flux in the unit of $\mathrm{10^{-17} erg/s/cm^2/Angstrom}$\\
  
 6 &
OII3728\_FLUX&
[O II]$\lambda3728\AA$ flux in the unit of $\mathrm{10^{-17} erg/s/cm^2/Angstrom}$\\                         
\hline

\end{tabular}

\label{tab:Rad Flux}
\end{table*}

\begin{table*}
\caption{Global emission line fluxes and their uncertainties in SFR\-$\mathrm{M_\star}$ bins. The structure and description of the table columns are laid out here, with the complete version available in the online journal.}

\begin{tabular}{rll}

Column &  Name & Description\\
\hline
1&
BIN\_ID &
Number of the bin according to fig\ref{fig:bins}, starting  from the lowest $\mathrm{M_\star}$ and SFR\\

2 &
Mass&
Median stellar masses estimate of all galaxies within each SFR--$\mathrm{M_\star}$ bin in the unit of $\mathrm{log(M_\star/M_\odot)}$\\

3 &
SFR&
Median star-formation rate of all galaxies within each SFR--$\mathrm{M_\star}$ bin based on MaNGA \textsc{DAP} catalog\\

4 &
OII3726\_FLUX&
[O II]$\lambda3726\AA$ flux in the unit of $\mathrm{10^{-17} erg/s/cm^2/Angstrom}$\\
  
5 &
OII3728\_FLUX&
[O II]$\lambda3728\AA$ flux in the unit of $\mathrm{10^{-17} erg/s/cm^2/Angstrom}$\\          
6 &
H93751\_FLUX&
H9$\lambda3751\AA$ (i.e. 9th emision line of Balmer series) flux in the unit of $\mathrm{10^{-17} erg/s/cm^2/Angstrom}$\\                  

\hline

\end{tabular}

\label{tab:Global fluxes}
\end{table*}

\begin{table*}
\caption{Radially binned inferred physical parameters, electron temperature, ionic abundance, $\mathrm{T_e}$ and stron-line metallicity for all the bins. The structure and description of the table columns are laid out here, with the complete version available in the online journal.}

\begin{tabular}{rll}

Column &  Name & Description\\
\hline
1&
BIN\_ID &
Number of the bin according to fig\ref{fig:bins}, starting  from the lowest $\mathrm{M_\star}$ and SFR\\

2 &
RAD\_BIN&
Indication of the integrated spectra within this radial range, normalized to $\mathrm{R_e}$\\

3 &
Mass&
Median stellar masses estimate of all galaxies within each SFR--$\mathrm{M_\star}$ bin in the unit of $\mathrm{log(M_\star/M_\odot)}$\\

4 &
SFR&
Median star-formation rate of all galaxies within each SFR--$\mathrm{M_\star}$ bin based on MaNGA \textsc{DAP} catalog\\

5 &
T\_NII&
Electron temperature of $\mathrm{N^+}$ in the unit of Kelvin \\
  
6 &
T\_NII\_ERR&
Electron temperature uncertainty of $\mathrm{N^+}$ in the unit of Kelvin\\          

\hline

\end{tabular}

\label{tab:Radially binned inferred}
\end{table*}

\begin{table*}
\caption{Global inferred physical parameters, including electron temperature, ionic abundances, $\mathrm{T_e}$-based metallicity, and strong-line metallicity for all bins. The structure and description of the table columns are laid out here, with the complete version available in the online journal.}

\begin{tabular}{rll}

Column &  Name & Description\\
\hline
1&
BIN\_ID &
Number of the bin according to fig\ref{fig:bins}, starting  from the lowest $\mathrm{M_\star}$ and SFR\\

2 &
Mass&
Median stellar masses estimate of all galaxies within each SFR--$\mathrm{M_\star}$ bin in the unit of $\mathrm{log(M_\star/M_\odot)}$\\

3 &
SFR&
Representative star-formation rate of all galaxies within each SFR--$\mathrm{M_\star}$ bin based on MaNGA \textsc{DAP} catalog\\

4 &
T\_NII&
Electron temperature of $\mathrm{N^+}$ in the unit of Kelvin \\
  
5 &
T\_NII\_ERR&
Electron temperature uncertainty of $\mathrm{N^+}$ in the unit of Kelvin\\          
6 &
T\_OII&
Electron temperature of $\mathrm{O^+}$ in the unit of Kelvin\\                  

\hline

\end{tabular}

\label{tab:Global inferred physical}
\end{table*}

\end{appendix}


%
\clearpage
\newpage
  \bibliographystyle{aa} 
  \bibliography{ref} 
%

\end{document}